\documentclass[twocolumn]{aastex6}
\usepackage{xcolor}
\usepackage{txfonts,graphicx,rotating}
\usepackage{afterpage}

\usepackage{natbib,twoopt}

\begin{document}

\defcitealias{Zhukovska:2008bw}{ZGT08}
\defcitealias{Zhukovska:2016jh}{ZDJ16}

\DeclareRobustCommand{\ion}[2]{\textup{#1\,\textsc{\lowercase{#2}}}}
\newcommand{\aav}{\ensuremath{ \langle a \rangle_3} }
\newcommand{\sih}{\ensuremath{ {\rm [Si_{gas}/H}]}}
\newcommand{\feh}{\ensuremath{ {\rm [Fe_{gas}/H}]}}
\newcommand{\HH}{\ensuremath{\rm H}}
\newcommand{\nh}{\ensuremath{n_{\rm H}}}
\newcommand{\si}{\ensuremath{\rm Si}}
\newcommand{\dx}{\ensuremath{\delta_{\rm X}}}
\def\sii{\ion{Si}{I}}
\def\siii{\ion{Si}{II}}
\def\ci{\ion{C}{I}}
\def\cii{\ion{C}{II}}
\newcommand{\mum}{\ensuremath{\rm \, \mu m}}
\newcommand{\Ms}{\ensuremath{\rm \,M_{\sun}}}
\newcommand{\Zs}{\ensuremath{\rm \,Z_{\sun}}}
\newcommand{\Mspc}{\ensuremath{\rm \,M_{\sun}\,pc^{-2}}}
\newcommand{\Msgyrpc}{\ensuremath{\rm \,M_{\sun}\,Gyr^{-1}\,pc^{-2}}}
\newcommand{\Msyr}{\ensuremath{\rm \,M_{\sun}\,yr^{-1}}}
\newcommand{\pyr}{\ensuremath{\rm \,yr^{-1}}}
\newcommand{\cms}{\ensuremath{\rm \,cm^{-2}}}
\newcommand{\cmc}{\ensuremath{\rm \,cm^{-3}}}
\newcommand{\gcmc}{\ensuremath{\rm g\,cm^{-3}}}
\newcommand{\kms}{\ensuremath{\rm \,km\,s^{-1}}}
\newcommand{\ddt}[1]{{{\rm d}\, {#1} \over{\rm d}\,t}}

\author{Svitlana~Zhukovska\altaffilmark{1}}
   \altaffiltext{1}{Max-Planck-Institut f\"ur Astrophysik, Karl-Schwarzschild-Str. 1, D-85741 Garching, Germany}

\author{Thomas~Henning\altaffilmark{2}}%
   \altaffiltext{2}{Max-Planck-Institut f\"ur Astronomie, K\"onigstuhl 17, D-69117 Heidelberg, Germany}

\author{Clare Dobbs\altaffilmark{3}}%
   \altaffiltext{3}{University of Exeter, Stocker Road, Exeter EX4 4QL, United Kingdom}

\shortauthors{Zhukovska, Henning and Dobbs}

\title{Iron and silicate dust growth in the Galactic interstellar medium: clues from element depletions}
 \shorttitle{Iron and silicate dust  growth in the ISM}

 \begin{abstract}
The interstellar abundances of refractory elements indicate a substantial depletion from the gas phase, that increases with gas density. Our recent model of dust evolution, based on hydrodynamic simulations of the lifecycle of giant molecular clouds (GMCs) proves that the observed trend for \sih{} is driven by a combination of dust growth by accretion in the cold diffuse interstellar medium (ISM) and efficient destruction by supernova (SN) shocks \citep{Zhukovska:2016jh}. With an analytic model of dust evolution, we demonstrate that even with optimistic assumptions for the dust input from stars and without destruction of grains by SNe it is impossible to match the observed \sih--\nh{} relation without growth in the ISM. We extend the framework developed in our previous work for silicates to include the evolution of iron grains and address a long-standing conundrum: ``Where is the interstellar iron?''. 
Much higher depletion of Fe in the warm neutral medium compared to Si is reproduced by the models, in which a large fraction of interstellar iron (70\%) is locked as inclusions in silicate grains, where it is protected from sputtering by SN shocks. The slope of the observed \feh--\nh{} relation is reproduced if the remaining depleted iron resides in a population of metallic iron nanoparticles with sizes in the range of 1--10\,nm. Enhanced collision rates due to the Coulomb focusing are important for both silicate and iron dust models to match the slopes of the observed depletion--density relations and the magnitudes of depletion at high gas density. 
  \end{abstract}


\maketitle

\section{Introduction}
Interstellar dust grains are a small, but a very important component of the interstellar medium of galaxies \citep{Dorschner:1995p7228}. Despite the importance of dust, its main formation route remains a controversial topic. For decades, it was supposed that all refractory grains form in stellar winds of evolved stars and cooling ejecta of supernovae (SNe). In such a formation scenario, however, it is difficult to explain the high depletion of refractory elements in the cold interstellar medium at the level of $ \gtrsim 95\%$  for Si and $ \gtrsim 99\%$  for Fe for two reasons. Firstly, the timescale of dust formation by stars is longer than the lifetime of grains against destruction in SN shocks (\citealt{Dwek:1980p490, Jones:1994p1037, Slavin:2015in}; but see also \citealt{2011A&A...530A..44J}). Secondly this scenario requires almost complete condensation of refractory elements in solid form in all matter returned to the ISM by stars. Low- and intermediate-mass stars produce dust efficiently only during the final thermal pulses on the asymptotic giant branch (AGB), while during the preceding evolution on the red giant branch and the first AGB phases stars loose mass efficiently with no or little dust condensation \citep{Gail:2009p512, McDonald:2011eoa}. Moreover, star cluster environment may affect dust condensation and reduce the net dust yield from massive AGB stars \citep{Zhukovska:2015hs}. Similarly, it is unlikely that all refractory elements ejected by core-collapse SNe are in solid form. Although recent observations reveal that grains can form efficiently in expanding SN ejecta  \citep{Barlow:2010cz, Gomez:2012fm, Indebetouw:2014bt, Bevan:2016gz, deLooze:2017cz}, the newly formed dust can be subsequently destroyed by reverse shocks, which substantially reduce the fraction of elements in dust in SN ejecta \citep{Bianchi:2007p2222, Silvia:2010p6576, Micelotta:2016jk}. 

These discrepancies are resolved in one-zone dust evolution models by assuming that the dust mass can grow in the ISM by gas-grain interactions \citep{Dwek:1980p490, Dwek:1998p67, Draine:2009p6616}. This mechanism of dust formation is also required by very recent dust evolution models based on hydrodynamical simulations of the ISM to match observed dust properties in galaxies \citep{Bekki:2015hn, McKinnon:2016ft, Zhukovska:2016jh, Aoyama:2017eq}. 

Direct evidence that metals go in and out of dust in the ISM comes from the observed variations of interstellar element depletions. For instance, the fractions of Si and Fe in the gas decreases by almost ten times between the warm and cold clouds in the galactic disk \citep{Savage:1996p486}. These differences are explained by simple dust evolution models based on idealised two- or three-phase models of the ISM in which grains are destroyed by SN shocks in the intercloud medium and accrete gas-phase elements in clouds; denser gas enriched with dust is exchanged with dust-poor gas during the interstellar matter cycle \citep{Draine:1990p495, Tielens:1998p7054, ODonnell:1997p683, Weingartner:1999p6573}. These models are, however, sensitive to the adopted scheme and timescales of mass exchange between ISM phases which are treated as free parameters. 

Using realistic hydrodynamic numerical simulations, \cite{Peters:2017bl} demonstrate that dust grains experience complex evolution in a multi-phase ISM characterised by large scatter in the phase exchange timescales. Their work underlines the importance of combining dust models with numerical simulations of the ISM, which incorporate a wide range of physical conditions. Moreover, dust evolution models based on simulations can address a strong observed correlation of element depletions with the average gas density on the line of sight \citep{Savage:1979ev}. Recently, \citealt{Zhukovska:2016jh}~(hereafter ZDJ16) have presented a new dust evolution model based on hydrodynamic simulation of giant molecular clouds in a Milky Way-like galaxy, whose resolution is sufficiently high to explore the dependence of  interstellar Si depletions on local physical conditions and for the first time compare them with the observational data. They find that a combination of selective accretion of Si on silicate grains in the cold neutral medium (CNM) and dust destruction by interstellar shocks in the diffuse gas are necessary to explain the observed relation between Si depletion and density. In this work, we demonstrate with an  analytical model that dust input from stellar sources cannot describe this relation even with optimistic assumptions for dust condensation efficiencies in stars. 

\begin{figure}[]
\includegraphics[scale=0.69]{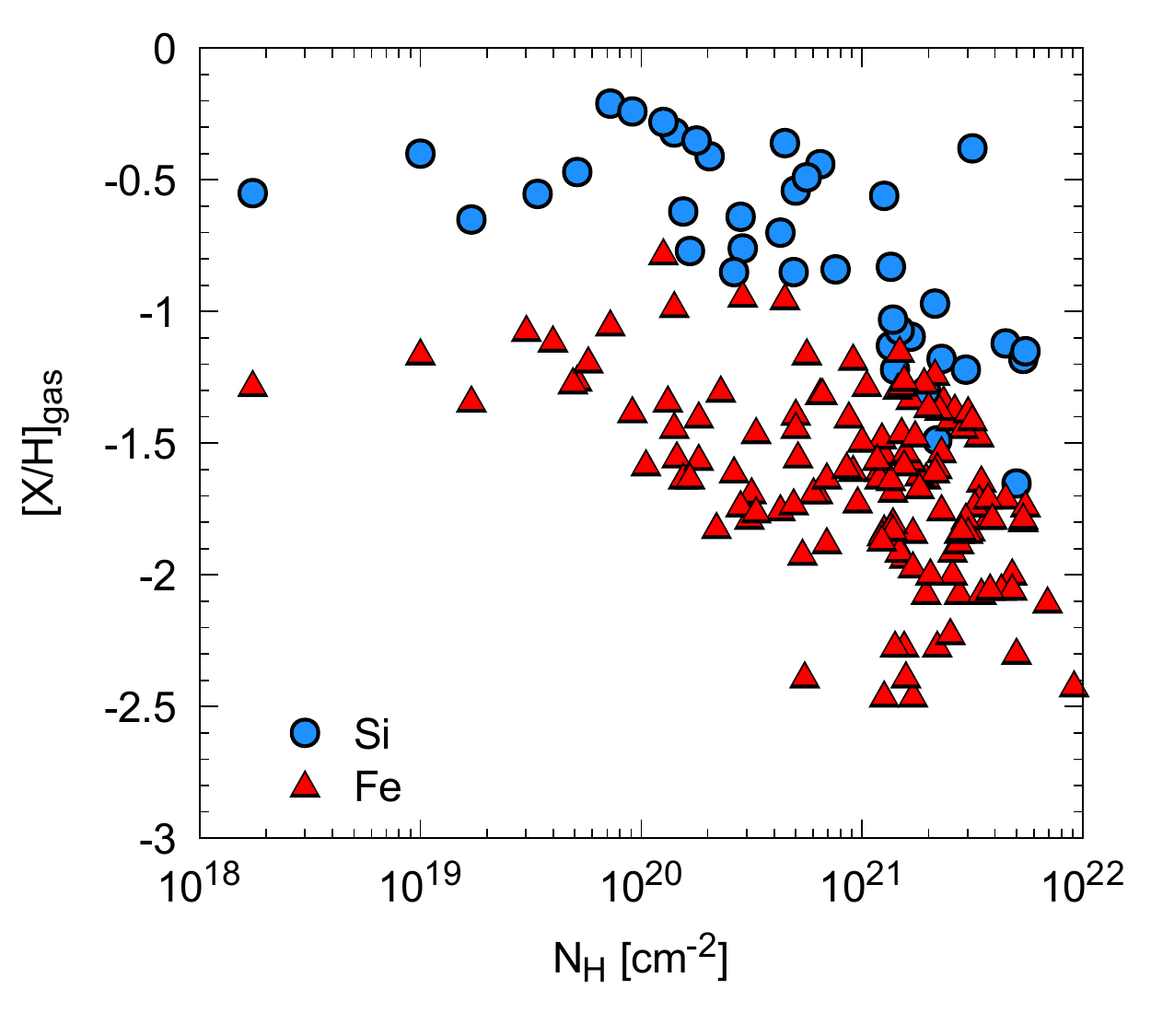}
\caption{Interstellar element abundances for Si and Fe (circles and triangles, respectively) as a function of hydrogen column density. Observational data are taken from work by \cite{Voshchinnikov:2010p6003}.} 
\label{fig:Column_Fe-Si}
\end{figure}

Iron represents a curious test case to investigate dust growth in the ISM with the evolution model developed in our previous work \citepalias{Zhukovska:2016jh}. In contrast to Si, it is more heavily depleted in the warm neutral medium (WNM) and, because of its comparable abundance to Si and Mg, can be an important component of interstellar dust. \cite{Dwek:2016kf} argues that, with the current observational data for iron dust input from various stellar sources, the observed Fe depletion can be explained only by accretion of gas-phase Fe on grains. The question ``In which solid form is the interstellar iron?'' poses a long-standing conundrum. The composition of interstellar silicates is not uniquely constrained by the observational data \citep[e.g.,][]{Zubko:2004p4116}, therefore the fraction of iron locked in silicates is unclear. \cite{Schalen:1965jq} suggested that the depleted cosmic Fe can be in the form of metallic iron, which unfortunately does not have extinction features for its unambiguous identification. \cite{Ossenkopf:1992p7790} include iron in both the silicate lattice and as small inclusions (pure iron and iron oxides) to the silicates to explain the large absorptivity of interstellar dust at the near-infrared wavelengths. Among other solid forms of iron considered in the literature are oxides \citep[][]{Henning:1995uo, Draine:2013haa}, metallic needles \citep{Dwek:2004fd}, iron and FeS inclusions in silicate grains \citep{Min:2007p476, Jones:2013gg},  an inner layer in multi-layered particles \citep{Voshchinnikov:2017fi}, free-flying iron nanoparticles \citep{Hensley:2017bc, Gioannini:2017jm}, and hydrogenated iron nanoparticles \citep{2017MNRAS.466L..14B}. Indeed, in-situ studies of interstellar dust grains demonstrate that silicate particles contain iron and that individual iron particles exist \citep{Westphal:2014gf, Altobelli:2016fw}.

In this work, we extend the framework introduced by \citetalias{Zhukovska:2016jh} to include iron dust evolution. We fix the ratio of carbon-to-silicate dust mass and postpone the evolution of carbonaceous dust to future. Observational constraints from interstellar element depletions are introduced in Sect.~\ref{sec:Observ}. A relation between Si gas-phase abundance and gas density is interpreted with an analytic model of dust evolution in Sect.~\ref{sec:SimpleMod}. We briefly summarise the hydrodynamic numerical simulations of the giant molecular clouds and  dust evolution model in Sect.~\ref{sec:SimDescr}. Section~\ref{sec:Si} introduces a model for the silicate grains and compares a synthetic relation between the average Si gas-phase abundance and gas density with the observed relation. In Sect.~\ref{sec:Fe}, we explore different models for the iron grain population to understand the reasons behind the higher depletion of the interstellar Fe abundance in the WNM compared to the Si depletions and the increasing trend of \feh{} with density. We consider the timescales of physical processes on the grain surface and the role of stochastic heating of small iron grains for the dust growth process in Sect.~\ref{sec:StochHeat}. Discussion of our results is given in Sect.~\ref{sec:Discussion}. Finally, our conclusions are presented in Sect.~\ref{sec:Conclusions}.

\begin{figure}[]
\includegraphics[scale=0.69]{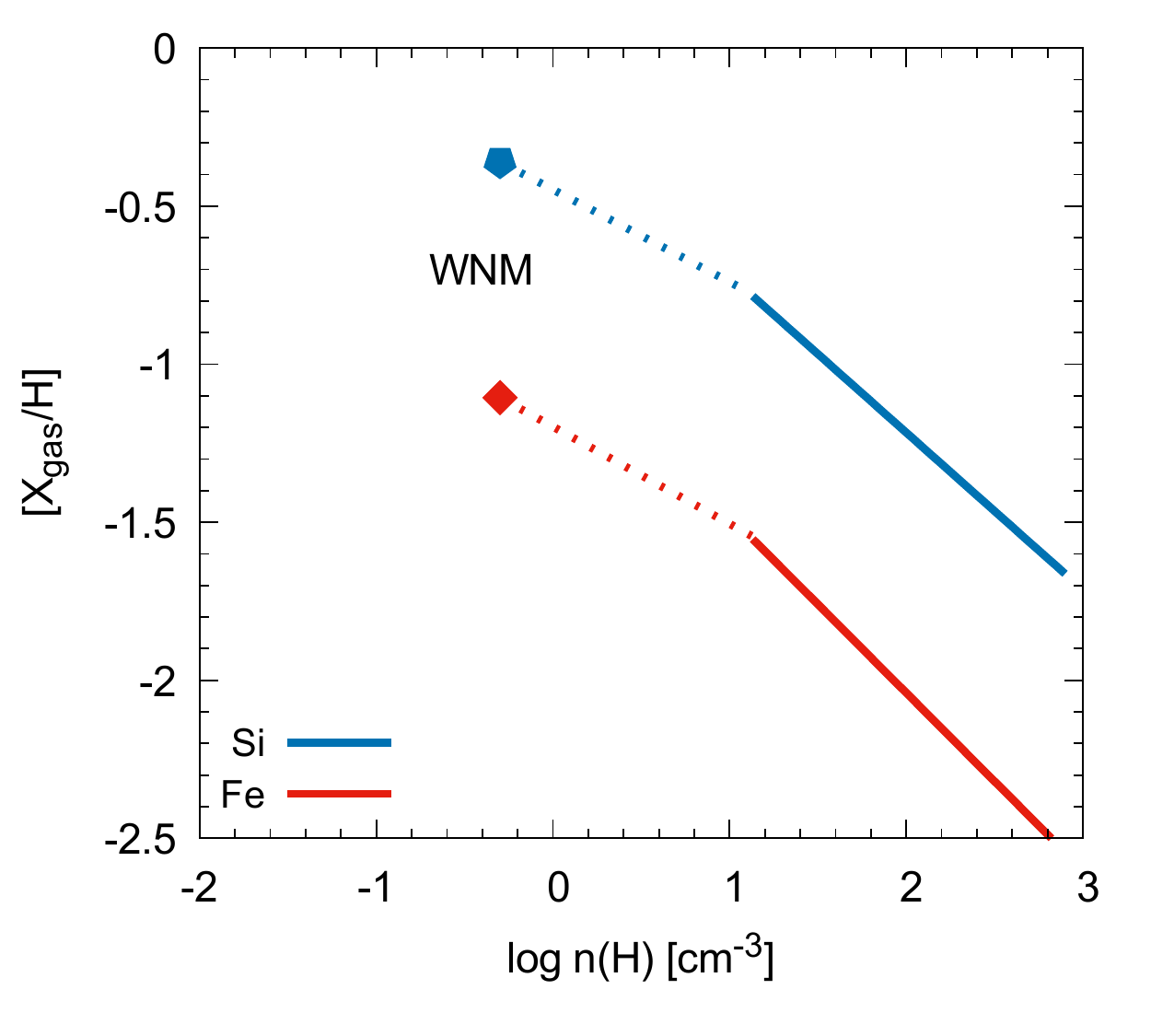}
\caption{Relations between element depletion and gas density for Si and Fe derived from the least-square linear fit to observational data (blue and red solid lines, respectively) described in Sect.~\ref{sec:Observ}. Pentagon and diamond show the depletion values for Si and Fe in the WNM. Dashed lines are interpolations between the observed values in the WNM and the linear fit relations.} 
\label{fig:Depl-dens-Obs}
\end{figure}

\section{Gas-phase Si and Fe depletions}\label{sec:Observ}
Evolutionary changes in dust composition can be studied by analysis of element depletions in the gas phase measured along various lines of sight, assuming that the elements missing from the gas phase are locked in grains. The logarithmic depletion of an element X in the ISM ${\rm [X/H]_{gas}}$ is determined as the gas-phase element abundance relative to the reference abundance, for which we adopt its abundance in the Sun:
\begin{equation}
{\rm [X/H]_{gas}} = \log\left( N({\rm X})/N(\HH) \right) - \log\left( N({\rm X})/N(\HH) \right)_{\sun},
\end{equation}
where $N({\rm X})$ refers to the  column density of an element X. The linear depletion of element X is, correspondingly,
\begin{equation}
 \dx = 10^{\rm [X/H]_{gas}}
\end{equation}

Abundances of main dust-forming elements in the gas phase decrease with hydrogen column density $N_{\rm H}$ ($N_{\rm H} = N(\ion{H}{I})+ N({\rm H_2})$) pointing to the increase of their abundances in dust. Figure~\ref{fig:Column_Fe-Si} demonstrates this  observational depletion trend for Si and Fe, which are assumed to be the key elements for silicate and iron dust, respectively. For $N_{\rm H}>5\times 10^{19}\,\cms$, \sih{} and \feh{} show similar behaviour, despite a 1 dex difference in their values. These changes in the element abundances are driven by dust evolution, which depends on the local gas density rather than on the column density measured observationally. The element depletions have been analysed in the literature as a function of the averaged gas density along the line of sight, but this quantity can be significantly lower than the actual volume density in a cloud. In our previous work \citepalias{Zhukovska:2016jh}, we derived a relation between \sih{} and the local gas volume density using fine structure excitations of neutral carbon from \cite{Jenkins:2011by} measured for 87 lines of sight. The volume density represents local physical conditions of interstellar gas, therefore the relation between \sih{} and the volume gas density is more suitable to constrain three-dimensional dust evolution models than the relation for the average density on the line of sight. We find that \sih{} correlates with the gas density as $\nh^{-0.5}$.

To derive the analogous relation between \feh{} and \nh, we combine the relation for Si given by Eq.~(20) from \citetalias{Zhukovska:2016jh} and Eq.~(10) from \cite{Jenkins:2009p2144} with coefficients for Si and Fe. The resulting equation is
\begin{equation}
\feh = -0.56 \log\left(\nh \right) - 0.72
\end{equation}
The slope of this relation $-0.56$ is slightly steeper than the value of $-0.5$ for Si indicating that Fe is more readily depleted from the gas phase than Si  (Fig.~\ref{fig:Depl-dens-Obs}).

The gas density range accessible through the $\ion{C}{I}$ fine structure lines is restricted to about 10--10$^3\cmc$. Since the element depletion is measured along the entire line of sight, the resulting ${\rm [X/H]_{gas}}$ value can be higher than the depletion in a cloud probed by $\ion{C}{I}$ because of the contribution of WNM on the same line of sight \citepalias[see][for details]{Zhukovska:2016jh}. Although the contamination from the WNM does not have a large effect on the coefficients in the $\sih - \nh$ relation, its magnitude can reach 0.1--0.2~dex for \sih{} (Jenkins, private communication). To account for this effect, the relations for Si and Fe are shifted downwards by 0.2 in Fig.~\ref{fig:Depl-dens-Obs}.

Figure~\ref{fig:Depl-dens-Obs} also includes the depletion levels in the WNM at a location for a typical density of the WNM $\nh=0.5\cmc$. The values $\feh=-1.11$ and $\sih=-0.36$ are calculated from Eq.~(10) in \cite{Jenkins:2009p2144} for $F_*=0.12$ recommended for the WNM. These values are slightly higher than the values $\feh=-1.22$ and $\sih=-0.43$ recommended for the warm disk by \cite{Savage:1996p486} based on a smaller data sample. Comparison of the depletion values for the WNM and the relations for \sih{} and \feh{} derived for $\nh\gtrsim10\cmc$ points to shallower slopes in the density range $0.5-10\cmc$.

\section{Analytic steady-state model}\label{sec:ModSimple}\label{sec:SimpleMod}
\subsection{Formulation}
In this section, we derive a conceptual model for the fraction of an element $X$ remaining in the gas $\dx$ as a function of the gas density aiming to explain the observational trends discussed in the proceeding section. 
We adopt a formulation of dust evolution in the solar neighbourhood \citep[][hereafter ZGT08]{Zhukovska:2008bw} and substantially simplify it assuming a steady-state situation for $\dx$. This assumption is appropriate to study the present-day element depletions in an evolved galaxy such as the Milky Way. This is an one-zone model, which considers the surface density of dust $\Sigma_{\rm d}$ averaged over the ISM phases within an annulus at the solar galactocentric radius. The average depletion is related to the dust surface density as  
\begin{equation}
\dx =1 -  \Sigma_{\rm d}/\Sigma_{\rm d, max},
\label{eq:delta-sigm}
\end{equation}
where $\Sigma_{\rm d,max}$ is the maximum surface density of dust, corresponding to the complete condensation of the key element X.
The evolution of $\Sigma_{\rm d}$ is described by the equation:
\begin{equation}
\ddt{\Sigma_{\rm d}} = R_{\rm d,inj} + R_{\rm d,gr} - \frac{\Sigma_{\rm d}}{\tau_{\rm SF}} -  \frac{\Sigma_{\rm d}}{\tau_{\rm d,SN}},
\label{eq:SigmRates}
\end{equation}
where $R_{\rm d,inj}$ is the rate of dust injection by stellar sources, $ R_{\rm d,gr}$ is the growth rate by gas-grain interactions in the ISM, the third and fourth terms on the r.h.s. are the rates of dust destruction by star formation and by SN blast waves, respectively, and $\tau_{\rm SF}$ and $\tau_{\rm SN}$ are corresponding timescales of these processes. The timescale of star formation is $\tau_{\rm SF}=\Sigma_{\rm ISM}/\psi$, $\Sigma_{\rm ISM}$ is the surface densities of dust and gas, $\psi$ is the star formation rate per unit area. The timescale of dust destruction by SN is provided by \citep{Dwek:1980p490}
\begin{equation}
\tau_{\rm d,SN} = \frac{\Sigma_{\rm ISM}}{f_{\rm SN} m_{\rm cl} R_{\rm SN}},
\label{eq:taud}
\end{equation}
where $f_{\rm SN}$ is the fraction of all SNe that destroy dust, $m_{\rm cl}$ is the mass of gas cleared of dust by a single SNe and $R_{\rm SN}$ is the current SN rate. $m_{\rm cl}$ is determined by the material and size distribution of dust grains \citep{Jones:1994p1037}. Assuming that only type II SNe destroy dust and approximating the SN rate by $R_{\rm SN} \approx \eta_{\rm SN} \psi$, where $\eta_{\rm SN} = 1/150\Ms$ is the number of SN per unit of stellar mass formed for the initial mass function (IMF) from \cite{2002Sci...295...82K}, we derive the relation between the timescales
\begin{equation}
\tau_{\rm d,SN} =  \frac{\tau_{\rm SF}}{f_{\rm SN} m_{\rm cl} \eta_{SN}}.
\label{eq:taud-tauSF}
\end{equation}
We introduce the effective destruction timescale
\begin{equation}
{1\over\tau_{\rm d,eff}}={1\over \tau_{\rm SF}} + {1\over \tau_{\rm SN}} = \frac{1+f_{\rm SN} m_{\rm cl} \eta_{SN}}{\tau_{\rm SF}}
\label{eq:taudeff}
\end{equation}
The rate of dust injection by stars is approximated as \citep[][]{ODonnell:1997p683, Weingartner:1999p6573}
\begin{equation}
R_{\rm d,inj} = \frac{(1-\delta_{\rm in}) \Sigma_{\rm d,max}}{\tau_{\rm in}},
\label{eq:RateInj}
\end{equation}
where $\delta_{\rm in}$ is the fraction of element X left in the gas in matter injected by stars and $\tau_{\rm in}$ is the timescale on which stars return matter to the ISM. An accurate expression for $R_{\rm d,inj} $ including its dependence on stellar mass and metallicity can be found in \citetalias{Zhukovska:2008bw}. 

We adopt a model of dust growth in the ISM proposed by \citetalias{Zhukovska:2008bw}, in which grains grow by accretion of elements in clouds characterised by the total mass fraction $X_{\rm cl}$ and lifetime $\tau_{\rm exch}$. It allows us to account for the finite time available for the dust growth by accretion before cloud dispersal. The dust growth rate is determined by the parameters of clouds as follows
\begin{equation}
R_{\rm d,gr} = \frac{f_{\rm X, ret}  \Sigma_{\rm d,max} - \Sigma_{\rm d}}{\tau_{\rm exch, eff}},
\label{eq:RateGr}
\end{equation}
where $\tau_{\rm exch, eff} = \tau_{\rm exch}(1-X_{\rm cl})/X_{\rm cl}$ is the effective timescale required for the entire ISM to cycle through the cloud phase, $f_{\rm X, ret}$ is the fraction of element $\rm X$ condensed in dust in the material returned from clouds. The corresponding depletion is $\delta_{\rm X, ret} = 1 - f_{\rm X, ret}$. We employ Eq.~(32) from \citetalias{Zhukovska:2008bw} for the time evolution of $f_{\rm X, ret}$ and, assuming that the depletion of gas forming clouds is not too different from the average depletion $\dx$, find the depletion at the end of the cloud lifetime ($t=\tau_{\rm exch}$): 
\begin{equation}
\delta_{\rm X, ret} = 1- f_{\rm X, ret} = \frac{\dx}{\dx+ (1-\dx)e^{\tau_{\rm exch}/\tau_{\rm gr}}  },
\label{eq:delta_ret}
\end{equation}
where $\tau_{\rm gr}$ the timescale of dust growth by accretion of the key species from the gas-phase given by
\begin{equation}
           \tau^{-1}_{\rm gr} = {3 \alpha_{\rm X} A_{\rm X,dust} m_{\rm amu}\over \rho_{\rm X,c} \nu_{\rm X,c}} \cdot {1\over \langle a_{\rm X} \rangle} \cdot \upsilon _{\rm X,th} n_{\rm H}   \cdot \epsilon_X \, ,
\label{eq:taugr}
\end{equation}
where 
$\alpha_{\rm X}$ is the sticking efficiency to the grain surface, 
$A_{\rm X,dust}$ is the atomic weight of one formula unit of the dust species,
$\rho_{\rm X,c} $ is the density and $\nu_{\rm X,c} $ is the number of atoms of the key element contained in the formula unit of the condensed phase, 
$\langle a_{\rm X}\rangle$ is the average grain radius,
$\upsilon _{\rm X,th}$ is the thermal speed of the growth species, 
and $\epsilon_{\rm X}$ is the element abundance of the key species.

\begin{figure*}[]
\includegraphics[scale=0.69, page=1]{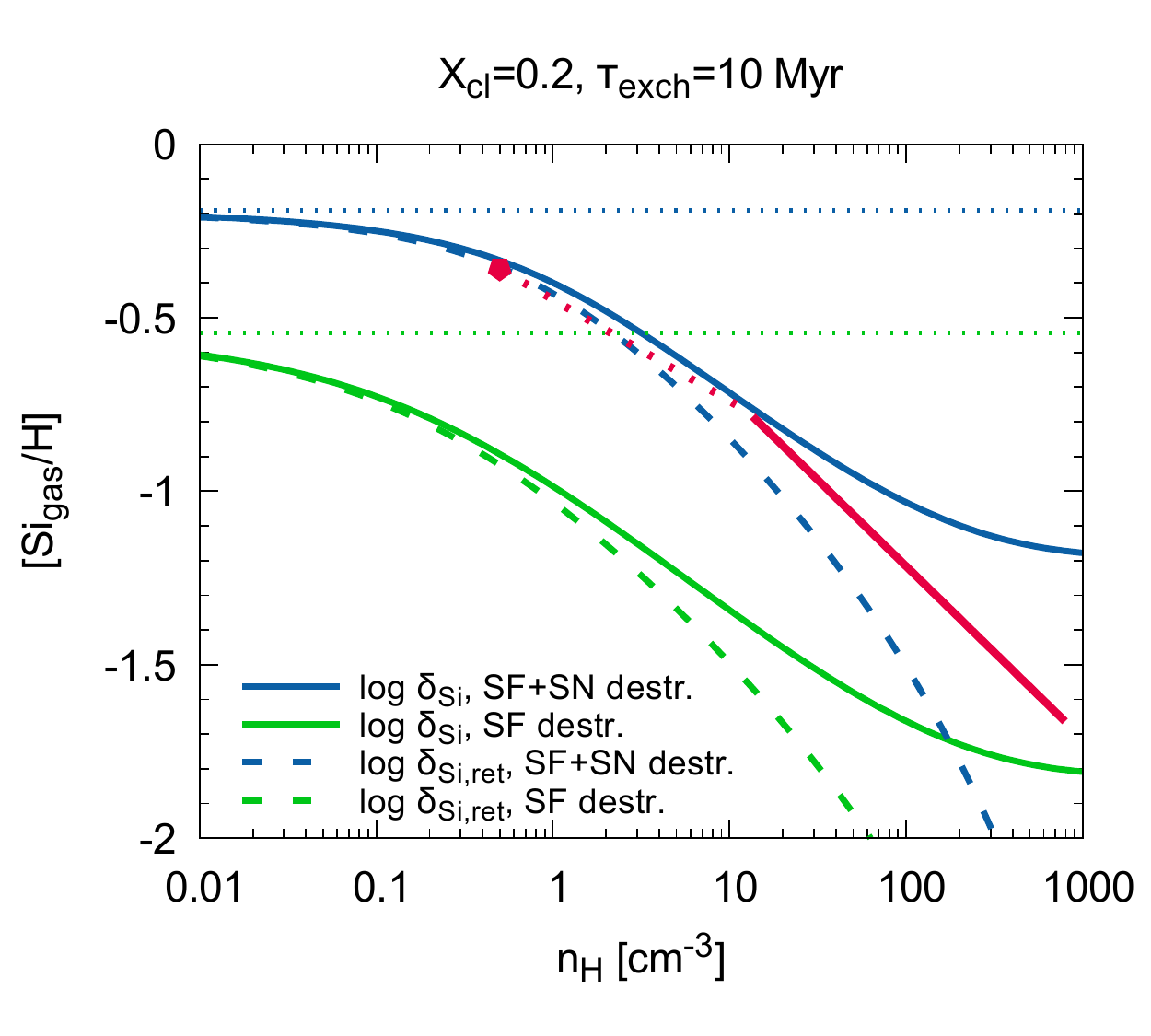}
\includegraphics[scale=0.69, page=2]{AnalytSolution_v4_deltfinal}
\caption{Si gas abundance $\sih \equiv \log \delta_{\rm Si}$ predicted by a simple steady-state model as a function of the gas density in clouds \nh{} (solid lines). Left and right panels show results for the lifetime of clouds of 10 and 30\,Myr, respectively. Blue  and green lines are for cases with and without dust destruction in SN shocks. Dashed lines demonstrate the final depletion $\log \delta_{\rm Si, ret}$ in the matter returned by clouds upon disruption. Depletion levels in models without dust growth in the ISM are shown with dotted lines. The observational data for Si are the same as in Fig.~\ref{fig:Depl-dens-Obs}.} 
\label{fig:AnalytSol}
\end{figure*}

We derive the expression for the evolution of $\dx$ by differentiating both sides of Eq.~(\ref{eq:delta-sigm})
\begin{equation}
\ddt{\dx} = 
 \frac{\dx - \delta_{\rm in}}{\tau_{\rm in}} + \frac{\dx - \delta_{\rm ret}}{\tau_{\rm exch,eff}}
- {(1-\dx)\over \tau_{\rm d,eff}}.
\label{eq:ddtdelta}
\end{equation}
In the derivation of Eq.~(\ref{eq:ddtdelta}), we assumed that ${\rm d}\Sigma_{\rm d,m}/{\rm dt} = \Sigma_{\rm d,m}/\tau_{\rm in}$ and used Eq.~(\ref{eq:SigmRates}) for ${\rm d}\Sigma_{\rm d}/{\rm dt}$, in which the expressions for dust destruction and production rates are re-written for $\dx$ with the use of Eq.~(\ref{eq:delta-sigm}). We also used Eq.~(\ref{eq:taudeff}) for the total destruction rate. 
In a steady state, ${\rm d}\dx/{\rm dt} = 0$ and Eq.~(\ref{eq:ddtdelta}) combined with Eq.~(\ref{eq:delta_ret}) yields a simple quadratic equation for $\dx$
%
\begin{eqnarray}
 \dx ( \tau_{\rm exch,eff}^{-1} +  \tau_{\rm in}^{-1}  + \tau_{\rm d,eff}^{-1} ) -    \nonumber \\ 
   - \frac{\dx  \tau_{\rm exch,eff}^{-1} }{\dx ( 1-e^{ \tau_{\rm exch}/\tau_{\rm gr}} )  + e^{\tau_{\rm exch}/\tau_{\rm gr}} }  
 -  \left( {\delta_{\rm in}\over \tau_{\rm in}} + {1\over \tau_{\rm d,eff}} \right)  = 0.
\label{eq:delta-main}
\end{eqnarray}
The solution of this equation determines $\dx$ as a function of density and temperature of clouds, for a fixed set of other model parameters. 
 Assuming that the ISM is an ideal gas in pressure equilibrium, we can replace the temperature dependence in Eq.~(\ref{eq:taugr}) by
\begin{equation}
T_{\rm gas} = T_0 \frac{n_0}{\nh},
\end{equation}
where we adopt the density $n_0=30 \cmc$ and temperature $T_0 = 100\,$K in the idealised CNM for the thermal pressure in the ISM. Then the growth timescale $\tau_{\rm gr}$ is 
\begin{equation}
\tau_{\rm gr} =  \frac{t_0}{ \nh^{0.5}},
\end{equation}
where $t_0$ is the constant determined by the dust properties given by Eq.~(\ref{eq:taugr}).

\subsection{Implications for \sih -- \nh{} relation}
In the following we compare predictions of the steady-state model for Si depletion $\delta_{\rm Si}(\nh)$ with the observed \sih--\nh{} relation (Sect.~\ref{sec:Observ}), which has been extensively studied with numerical simulations by \citetalias{Zhukovska:2016jh}.  

Following \citetalias{Zhukovska:2008bw}, we adopt a fixed power law distribution of grain sizes with the lower and upper size limits of $0.005$ and $0.25\mum$, respectively, and the power of $-3.5$ \citep[][hereafter called MRN]{Mathis:1977p750}. The model parameters in the solar neighbourhood are also taken from \citetalias{Zhukovska:2008bw}, $f_{\rm SN}=0.35$, $\psi=4 \Ms \, pc^{-2}\, Gyr^{-1}$, $\Sigma_{\rm ISM} = 10 \rm \Ms\, pc^{-2}$, $X_{\rm cl}=0.2$  and, 
for silicate dust, $t_0=87 \rm \,Myr$ and $m_{\rm cl}=1500\Ms$. 
The dust destruction timescales are then $\tau_{\rm SF}=2.5\,$Gyr, $\tau_{\rm d,SN}= 0.7\,$Gyr and $\tau_{\rm d,eff}= 0.55\,$Gyr. For the timescale of dust injection by stars $ \tau_{\rm in}$, we adopt an optimistic value of 1~Gyr. Full time-dependent dust evolution models predict much higher values of $ \tau_{\rm in}$ exceeding 9\,Gyr for the present day (\citealt{Dwek:1998p67}, \citetalias{Zhukovska:2008bw}).

Figure~\ref{fig:AnalytSol} illustrates solutions of Eq.~(\ref{eq:delta-main}) for $\log \delta_{\rm Si} \equiv \sih$ for models with and without dust destruction by SN shocks. Grains are destroyed in the process of star formation in all models. We assume complete condensation in stars ($\delta_{\rm in}=0$) to evaluate the role of stars for \sih{} distribution in the case of their highest possible contribution. We also show the final depletions $\log \delta_{\rm Si, ret}$ given by Eq.~(\ref{eq:delta_ret}) in the matter returned by clouds upon their disruption.

At low gas densities, the dust input rate from stars is higher than the growth rate in the ISM, because the dust growth is inefficient ($\delta_{\rm X, ret} \simeq \dx$) for $\tau_{\rm gr} \gg \tau_{\rm exch}$. The Si depletion is therefore determined by the balance between dust input from stars and destruction, which does not depend on $n_{\HH}$ under our assumptions. Therefore \sih--\nh{} relation flattens at lower densities. Despite the assumption of the complete condensation in stars and the short timescale of dust formation by stars, the lowest value of \sih{} provided by stellar sources in models with and without dust destruction in SN shocks are $-0.2$ and $-0.55$, respectively (Fig.~\ref{fig:AnalytSol}). These values are well above the observed depletions in the CNM.

We also consider a model without dust growth in the ISM, in which we include the dependence of $m_{\rm cl}$ on ambient density considered in Sect.~\ref{sec:A1} of Appendix. In this case, \sih{} level is determined by the balance between the stardust input and dust destruction in the ISM. Because the $m_{\rm cl}$ decreases with ambient density, this model also produces a negative slope, but it is too shallow compared to observations. The lowest \sih{} value of $-0.3$ is higher than measurements in the WNM and CNM.

With the increase of the cloud density, \sih{} decreases due to higher rate  dust growth by accretion in clouds. Models with dust  destruction only by star formation overpredict Si depletions for the entire density range. Models with dust destruction by SNe match the observed depletion in WNM and enclose the linear fit to the observational data available for $\nh > 10\cmc$ between the steady-state values of $\log \delta_{\rm Si}$ in the ISM and the maximum depletion reached in clouds $\log \delta_{\rm Si, ret}$. This conclusion holds for the current assumption of the complete condensation in stars and a short timescale $\tau_{\rm in}$. If stars are less efficient in dust formation, the model requires a lower efficiency of dust destruction in shocks (lower $m_{\rm cl}$).
 
The shape of the model \sih--\nh{} relation depends on the residence time of grains in clouds, i.e. the lifetime of clouds in our model. For simplicity, we adopt fixed $\tau_{\rm exch}$ values of 10 and 30\,Myr for short and long lifetimes of clouds, respectively. The difference between $\log \delta_{\rm Si}$ and $\log \delta_{\rm Si, ret}$ for $\tau_{\rm exch}=10$\,Myr corresponds to the scatter of observational data of up to 1~dex \citepalias[See Fig.~2 in][]{Zhukovska:2016jh}, but appears too large for $\tau_{\rm exch} =30$\,Myr.  In the dense gas, the growth timescale is shorter than the lifetime of clouds for both adopted values of $\tau_{\rm exch}$, therefore grains rapidly accrete available metals from the gas long before the matter is released from clouds. A longer $\tau_{\rm exch}$ means a delayed enrichment of the ISM with highly depleted gas from clouds, which results in the flattening of the \sih--\nh{} relation for $\nh\gtrsim50\cmc$. The observational relation favour the shorter lifetime of clouds of 10\,Myr. This lower value is also in better agreement with the residence times in the molecular and cold phases inferred from numerical hydrodynamic simulations by \cite{Peters:2017bl}.

The simple steady-state model for \sih{} explains the observed depletion trend with density as the result of the density dependence of the timescale of dust growth in the ISM. The $\sim$1~dex scatter in the observational data is explained by the gradual depletion of Si  in the process of dust growth, with the lowest depletions limited by the residence times of grains in clouds.

\section{Numerical simulations}\label{sec:ModSimul}
 In order to model the element depletions self-consistently and account for the dependence of dust evolution on local physical conditions, one has to solve a time-dependent problem including complex dynamical evolution of gas from numerical hydrodynamic simulations with sufficiently high resolution. In the following we discuss the  $\sih - \nh$ relation predicted with such a dust evolution model in an evolving multi-phase ISM.
 
\subsection{Description of the model}\label{sec:SimDescr}
The model presented here is a combination of the hydrodynamical simulations of the ISM in a Milky Way-like disk and a dust evolution model including dust destruction by SNe and dust growth by accretion of gas-phase metals in the ISM. The detailed description of the model can be found in \citetalias{Zhukovska:2016jh}.  It utilises a hydrodynamic SPH simulation of destruction and formation of giant molecular clouds (GMCs) in a spiral Milky Way-like galaxy described in \cite{Dobbs:2013hb}. The simulation is performed using the SPH code sphNG \citep{Benz:1990ct, Bate:1995wm, Price:2007jo}. 
 There is a total of 8 million gas particles, and each one has a mass $m_{\rm SPH}=312.5\Ms$. The gas is situated in a disk with radius of $10\,$kpc and the average surface density of the gas is $\Sigma_{\rm gas}=8\Mspc$. The total simulation time of 270\,Myr permits the system to reach steady state between destruction and growth of dust, which strongly correlates with the lifecycle of GMCs. 
Dust is modelled in a post-processing scheme using histories of the physical conditions from hydrodynamic simulations. Assuming perfect coupling between dust and gas, we model evolution of $\dx$ in each SPH gas particle. In this paper, we update the model for element Si and extend it to include Fe.

The dust model accounts for the dependence of dust growth by gas-grain collisions on local gas temperature and density.  \citetalias{Zhukovska:2016jh} found that growth of silicate dust is so efficient that it has to be suppressed in the warm medium, therefore it is limited to gas temperatures below 300\,K in this work. Additionally, enhanced collision rates in the cold neutral medium (CNM) due to Coulomb focusing are important to reproduce the observed \sih{}--\nh{} relation.

The grain size distribution is crucial for dust growth in the ISM as it determines the total grain surface area. The accretion timescale is proportional to the average grain radius  
\begin{equation}
 \aav = \langle a^3 \rangle/ \langle a^2 \rangle,
\label{eq:a3normal}
\end{equation}
where $\langle a^l \rangle \sim \int {{\rm d}  n_{\rm gr}(a)} /  { {\rm d}a}\, a^l {\rm d}a$ is the $l$th moment of the grain size distribution, $ {\rm d}  n_{\rm gr}(a) $ is the number of grains with radii from $a$ to $a+{\rm d}a$. The mean grain radius modified by the effect of Coulomb focusing is 
\begin{equation}
	\langle a^{\rm m} \rangle_3 = \langle a^3 \rangle/ \int  {\rm d} n_{\rm gr}(a) / {\rm d}a \, D(a)  a^2  {\rm d}a,
	\label{eq:a3}
\end{equation}
where the enhancement factor $D(a)$ accounts for the change in the cross section of an interaction between ion and grain \citep{Weingartner:1999p6573}. For neutral particles in MCs, $D(a)=1$. For simplicity, we adopt the MRN grain size distribution ${\rm d} n_{\rm gr}(a)/{\rm d}a \sim a^{-3.5}$ \citep[][]{Mathis:1977p750} with the lower and the upper limits for the grain sizes $a_{\min}$ and $a_{\max}$ that differ for iron and silicate grains.

The rate of dust destruction consists of two terms: destruction by supernova feedback in GMCs included in simulations and, additionally, destruction in the diffuse ISM by single SNe. The latter is implemented in a simplified way for all gas particles with $\nh<1\cmc$ assuming a constant destruction timescale of $\tau_{\rm dest}^{\rm diff}$, which can be expressed as
\begin{equation}
\tau_{\rm dest}^{\rm diff} = \frac{f_{\rm diff} \Sigma_{\rm gas}}{m_{\rm cl} R^{\rm diff}_{\rm SN}},
\label{eq:taudestr}
\end{equation} 
where $f_{\rm diff}$ is the mass fraction of the diffuse gas determined from the simulations and $R^{\rm diff}_{\rm SN}$ is the rate of SN that explode in the diffuse gas. We assume $R^{\rm diff}_{\rm SN}=8\times 10^{-12} \rm pc^{-2} yr^{-1}$. Numerically, we have $f_{\rm diff}(\nh<1\cmc)=0.35$. Note that the destruction timescale of all dust in the diffuse medium $\tau_{\rm dest}^{\rm diff}$ is shorter than the total destruction timescale in the ISM given by Eq.~(\ref{eq:taud}). 

The dependence of $m_{\rm cl}$ on density \nh{} is considered in Sect.~\ref{sec:A1} in Appendix. To examine how this dependence affects our results, we run test simulations of dust evolution using the fitting formula for $m_{\rm cl}(\nh)$ given by Eq.~(\ref{eq:A1}) in the expression for dust destruction rates. We find that the differences compared to the models with the fixed $m_{\rm cl}$ constitute less than 1\%. We therefore adopt a fixed value of $m_{\rm cl}$ derived for a typical density in the diffuse gas of $0.1\cmc$.

We name the models ECMRN\textit{x}nm following \citetalias{Zhukovska:2016jh}, where $x$ is the minimum grain size in nm and MRN indicates that all models assume an MRN grain size distribution. ``E'' and ``C'' in the name denote the enhanced collision rates due to Coulomb interactions and additional destruction in the diffuse ISM, respectively.

In the following, we apply the dust evolution model to determine the characteristics of grain populations that are responsible for the observed differences in the \feh{} and \sih{} depletion levels in the ISM.

\subsection{Silicon}\label{sec:Si}
Here we briefly summarise the model for silicate dust and refer to \citetalias{Zhukovska:2016jh} for details.
We model silicate evolution as traced by the Si abundance. To convert the Si abundance to the dust mass, we adopt amorphous silicates with olivine stoichiometric ratios. The dust evolution calculations start with homogeneous dust distribution with $\sih=-0.5$ or 68\% of Si in dust. The total Si abundance $\epsilon_{\rm Si}=3.548 \times 10^{-5}$  \citep{Asplund:2009p6572}. 

We estimate the destruction timescale in the diffuse gas from Eq.~(\ref{eq:taudestr}) assuming a fixed value $m_{\rm cl}=1500\,\Ms$ \citep{Zhukovska:2008p7215}, which results in the timescale of destruction in the diffuse gas $\tau_{\rm dest}^{\rm diff}=220$\,Myr. The timescale of silicate dust destruction in the entire ISM equates to about 350\,Myr in the steady state between destruction and growth of dust in the ISM. The slope of the observed $\sih-\nh$ relation is best reproduced with $a_{\min}=4\,$nm and $a_{\max}=250$\,nm. These grain size limits are in excellent agreement with the classical MRN size distribution. The average grain radii are $\aav=32$ and $\langle a^{\rm m} \rangle_3=5.6$\,nm, without and with the account of Coulomb focusing, respectively. For a typical density and gas temperature in the CNM (\nh=30\cmc{} and $T_{\rm gas}=100$\,K), the corresponding timescales of silicate dust growth are 15 and 1\,Myr. 

\begin{figure}[t]
\includegraphics[scale=0.69, page=1]{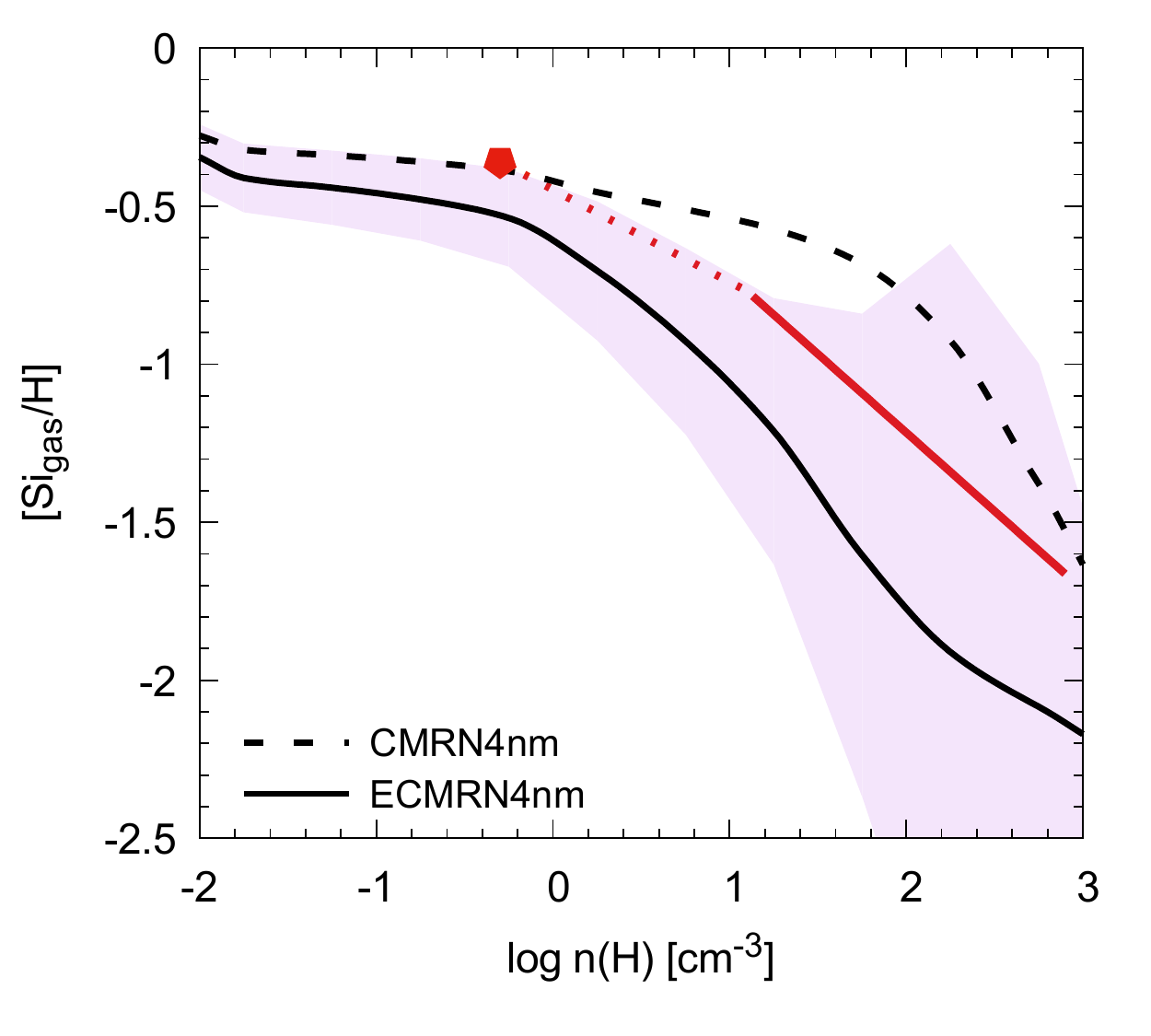}
\caption{Relation between mean Si depletion and gas density derived from the final spatial distribution of gas-phase Si abundance in simulations and its standard deviation (ECMRN4nm, solid black line and a shadow area around it, respectively). The relation for the model without enhanced collision rates (CMRN4nm, dashed line) is shown for comparison. Observational data are the same as in Fig.~\ref{fig:AnalytSol}.}
\label{fig:SiH-dens-rel}
\end{figure}

Figure~\ref{fig:SiH-dens-rel} shows the relation between the mean \sih{} and \nh{} predicted by the model and the relation inferred from observations in the local Milky Way. It is computed for the final distribution of gas and dust within a ring with galactic radii from 6 to 9\,kpc that represents the conditions similar to the solar neighbourhood. The mean \sih{} is derived as described in \citetalias{Zhukovska:2016jh} using mass-weighted probability density functions (PDFs). In this work, the PDFs are constructed for the linear depletion instead of the logarithmic depletion, therefore the slopes of the relation shown in Fig.~\ref{fig:SiH-dens-rel} are somewhat shallower than those presented by  \citetalias{Zhukovska:2016jh} in their Fig.~8. 
 
 
\subsection{Iron}\label{sec:Fe}
\subsubsection{Model parameters}
The iron element abundance in the Sun is similar to that of Si. Therefore, \feh{} values are expected to be similar to that of \sih{}, if Fe is also accreted on the silicate grains and Si and Fe have similar sputtering rates in SN shocks. \citetalias{Zhukovska:2016jh} found that the depletion levels in the WNM are determined by the destruction rates, while the depletions at high densities are controlled by the accretion rate. In order to reproduce the heavy depletion of interstellar Fe in the cold medium, we need a higher accretion rate of gas-phase Fe than that on the silicate grain surfaces. It can be provided if Fe is accreted on a population of small grains, as has been shown for another heavily depleted element Ti \citep{Weingartner:1999p6573}. 

 \cite{Hensley:2017bc} recently demonstrated that even a large population of metallic iron nanoparticles with sizes down to $4.5\, \AA$ can be a viable component of the interstellar grains. We therefore include a population of metallic nanoparticles that accrete Fe from the gas phase and are destroyed by interstellar shocks. The size distribution of such grains is unknown. We assume a power law size distribution with the lower size limit $a_{\min}=1$\,nm and the upper size limit $a_{\max}=10$\,nm, resulting in the mean particle size \aav=3.2\,nm. This is 10 times smaller than for silicate grains.

The fraction of iron in nanoparticles required to match $\feh=-1.11$ in the WNM depends critically on the rate of destruction of iron dust. The latter is determined by the efficiency of metallic iron destruction as a function of shock velocity through Eq.~(\ref{eq:MassCleared}) for $m_{\rm cl}$. Unfortunately, extensive theoretical studies of grain destruction in SN blast waves  do not provide such data for metallic iron \citep{Jones:1994p1037, Jones:1996p6593}. Because small grains are sputtered in interstellar shocks more efficiently than large grains, we adopt the same value of 1500\Ms{} as for silicates, as a lower limit for $m_{\rm cl}$. As we illustrate in the following section, even this low value of $m_{\rm cl}$ results in too high destruction rates and, respectively, too high \feh{} values in the WNM, if all solid Fe is placed in nanoparticles. 

We consider two solutions to mitigate this discrepancy. The first solution assumes a two-component iron dust model, in which Fe resides in metallic nanoparticles and as metallic inclusions in other grains, for example, Mg-rich silicate grains. Existence of metallic  inclusions in Mg-rich silicates is supported by X-ray observations in the CNM \citep{2012A&A...539A..32C}. Such inclusions may form when iron nanoparticles stick on the silicate grains and are subsequently covered by silicate layers in the process of accretion and, at higher gas densities, coagulation of silicate grains. Non-detection of iron dust in SN type Ia, the major source of iron in the Universe, suggests an interstellar origin of most of the solid iron as opposed to thermal condensation in the cooling circumstellar shells and SN ejecta \citep[see also][]{Dwek:2016kf}. This is corroborated by recent microgravity experiments \citep{Kimura:2017jc} that find a low condensation efficiency of metallic iron at high temperatures.

We make a key assumption that the metallic inclusions are protected inside of these grains from rapid destruction by interstellar shocks. The inclusions will be eventually exposed to the surface as the layers covering them are sputtered away. In this way, destruction may be the process that releases iron nanoparticles back to the diffuse gas and thus increases the surface area available for the subsequent accretion in the cold phase. A certain fraction of iron inclusions should be also destroyed, but we presently neglect this destruction term, assuming that it is small. We leave more detailed modelling of iron cycle in and out of silicate grains for future studies. 

\begin{figure}[t]
\includegraphics[scale=0.69]{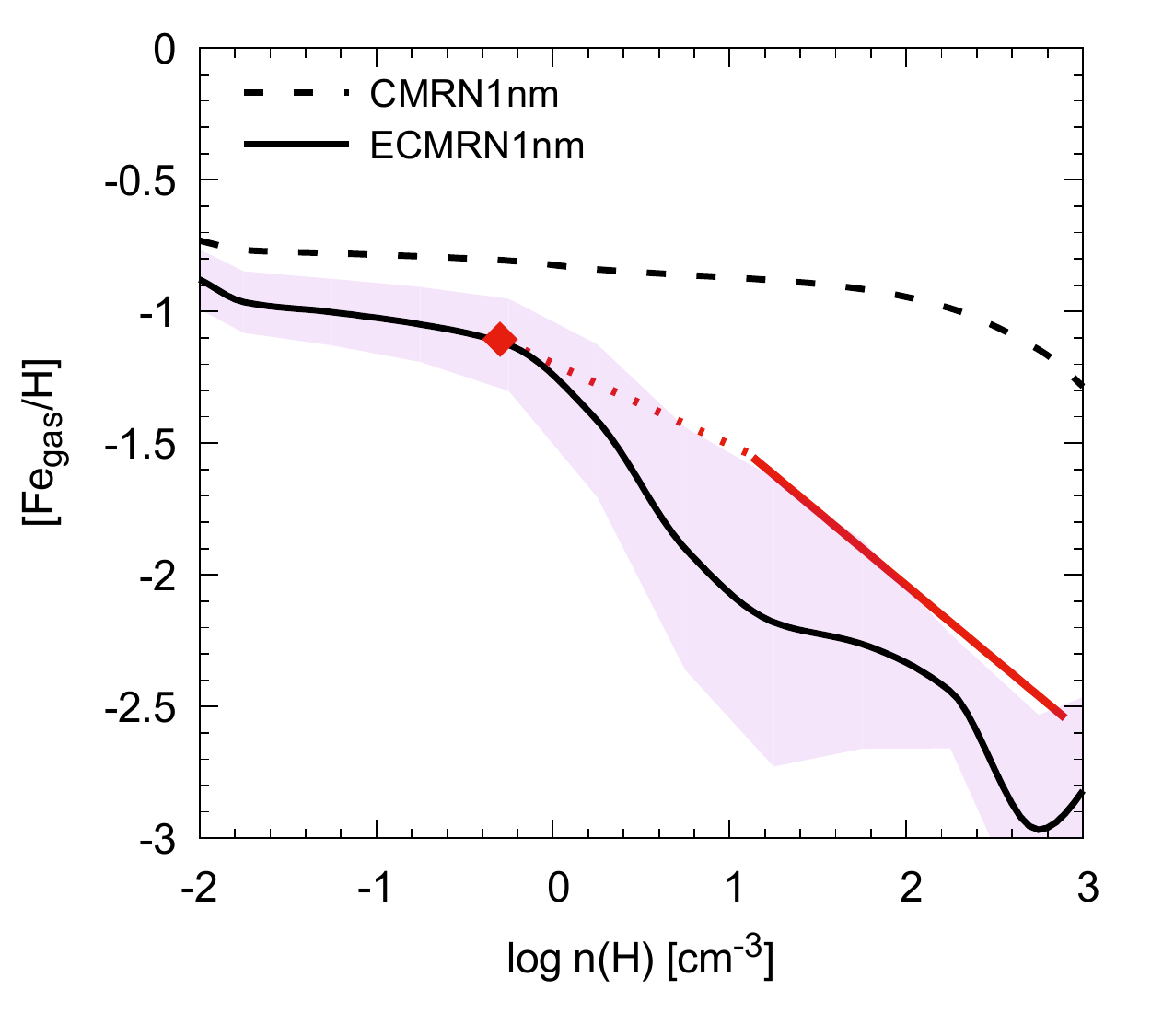}
\caption{The same as in Fig.~\ref{fig:SiH-dens-rel} for iron gas-phase abundance \feh{} derived for reference model ECMRN1nm. The relation for the model without Coulomb focusing (CMRN1nm, dashed line) is shown for comparison. Observational data for \feh{} are the same as in Fig.~\ref{fig:AnalytSol}.} 
\label{fig:FeH-dens-rel}
\end{figure}

A fixed fraction of iron locked in silicates $\eta_{\rm Fe}$ is adopted here, for which we take the value of 0.7 adopted in a new model of interstellar silicates by \cite{Jones:2013gg}. Although the destruction timescale of free-flying nanoparticles is short ($\tau_{\rm dest}^{\rm diff}=220$\,Myr), the average timescale of destruction of iron dust $\tau_{\rm dest}^{\rm ISM} = 900$\,Myr is longer than that for silicates because of the second iron dust component, metallic inclusions. This is our reference model.

Another scenario, in which all solid iron resides in free-flying metallic nanoparticles, requires the destruction of these particles to be less efficient than that of silicates to match the depletion in the WNM. It is hard to justify this scenario given that postshock velocities of iron grains are 15\%-35\% higher than equivalent silicate grain velocities \citep{Jones:1994p1037}. We examine the value of $m_{\rm cl}$ needed to fit $\feh$ in the WNM for the case when all solid iron resides in nanoclusters.

In order to compute the mean grain radius modified by Coulomb focusing $\langle a^{\rm m} \rangle_3$, we adopt the charge distribution functions for iron nanoparticles computed by \cite{Hensley:2017bc}. As for silicates, we consider the dust growth only at temperatures below 300\,K and compute the enhancement coefficient $D(a)$ for the grain charge distribution in the CNM. The majority of iron nanoparticles in the CNM have zero charge. 
 However, a substantial fraction of iron nanoparticles has the negative charge of $-1$: 0.36, 0.31, and 0.20 for the particle sizes of 1, 5, and 10\,nm, respectively. We find that the collision rate between gas-phase iron and metallic nanoparticles is significantly enhanced by Coulomb focusing on these negatively charged grains, resulting in $\langle a^{\rm m} \rangle_3=0.038$\,nm. This process decreases the timescale of growth of metallic grains in the CNM by a factor of 84, from 14\,Myr to 0.16\,Myr, respectively, estimated for typical conditions in the CNM (\nh{}=30\cmc, $T_{\rm gas}=100$\,K).

The electron field emission from negatively charged grains can be important for smaller grain sizes considered in this work, since the critical grain charge for this process decreases with the grain size as $Z_{\rm d(cr)} \simeq -210 \left( a/100\AA \right)^2$ \citep{Tielens:2005p6024}. For the lower size limit adopted in this work, the critical grain charge is $Z_{\rm d(cr)} = -2.1$. The negative grain charge attained in the CNM is $-1$, which is higher than $Z_{\rm d(cr)}$ and we can safely assume that electron field emission is not important in calculations of the accretion rates.

The simulations of iron dust evolution begin with the homogeneous iron abundance distribution with $\feh=-1.25$ corresponding to the Fe depletion in diffuse gas. As the dust evolves, the average \feh{} distribution more strongly correlates with the gas density due to accretion of gas-phase Fe on grains until the system reaches a steady state between destruction and accretion in the ISM.

\begin{figure}[t]
\includegraphics[scale=0.69, page=1]{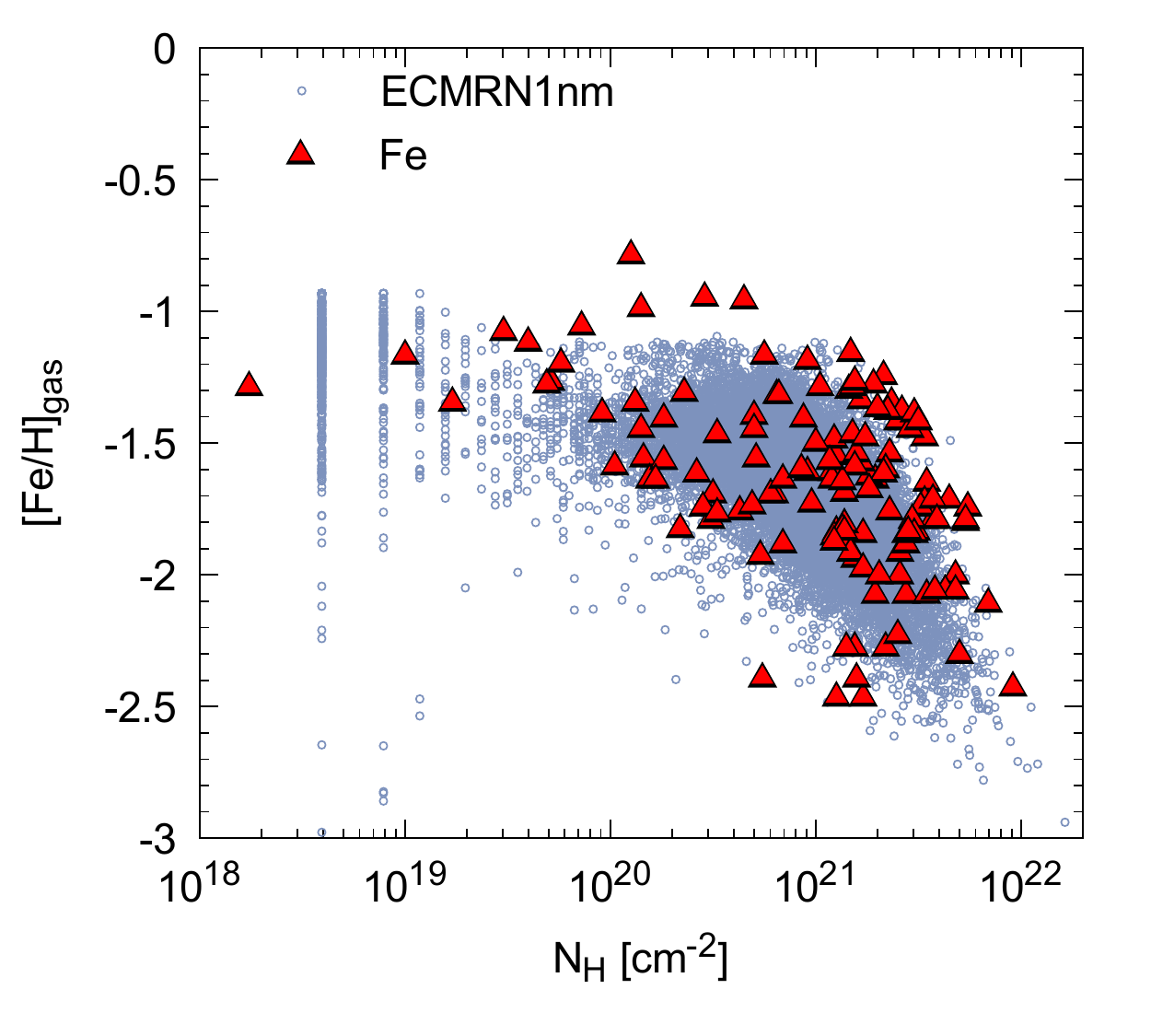}
\caption{Interstellar Fe abundances as a function of hydrogen column density for reference model for iron dust evolution ECMRN1nm (circles) calculated within 100\,pc sized cells in the direction perpendicular to the disk plane. Triangles show the values from observations \citep{Voshchinnikov:2010p6003}. }
\label{fig:Column-FeH-dens-rel}
\end{figure}

\subsubsection{Variation of \feh{} with gas density}
Figure~\ref{fig:FeH-dens-rel} shows the synthetic $\feh-\nh$ relations computed for the final spatial iron dust distribution for the reference model ECMRN1nm. Model CMRN1nm without Coulomb focusing is shown for comparison. The adopted size distribution for iron nanoparticles in model ECMRN1nm with the Coulomb focusing provides a high accretion rate of gas-phase iron and, correspondingly, a steep slope similar to the observed $\feh-\nh$ relation. The effective destruction timescale $\tau_{\rm dest}^{\rm ISM} = 900\,$Myr, on the other hand, sets the \feh{} level in the low density region. We did not vary the uncertain size distribution of nanoparticles to improve the fit to the observed relation.

Figure~\ref{fig:Column-FeH-dens-rel} shows the interstellar iron abundance in the gas phase \feh{} as a function of hydrogen column density of the gas $N_{\HH}$ from simulations for reference model ECMRN1nm and that from observational data from \cite{Voshchinnikov:2010p6003}. \feh{} and $N_{\HH}$ are calculated within 100\,pc sized cells in the directions perpendicular to the disk plane. The model can reproduce well the flattening of \feh{} at low $N_{\HH}$ and the decreasing trend for $N_{\HH}>10^{20}\cms$. Thus, our simulations  agree with the observational data for \feh{} as a function of both column hydrogen density characterising integrated properties of the ISM on the line of sight and the local density \nh. Large scatter in the predicted \feh{} values reflects different complex dynamical histories of the gas with the same $N_{\HH}$.

\begin{figure}[t]
\includegraphics[scale=0.69]{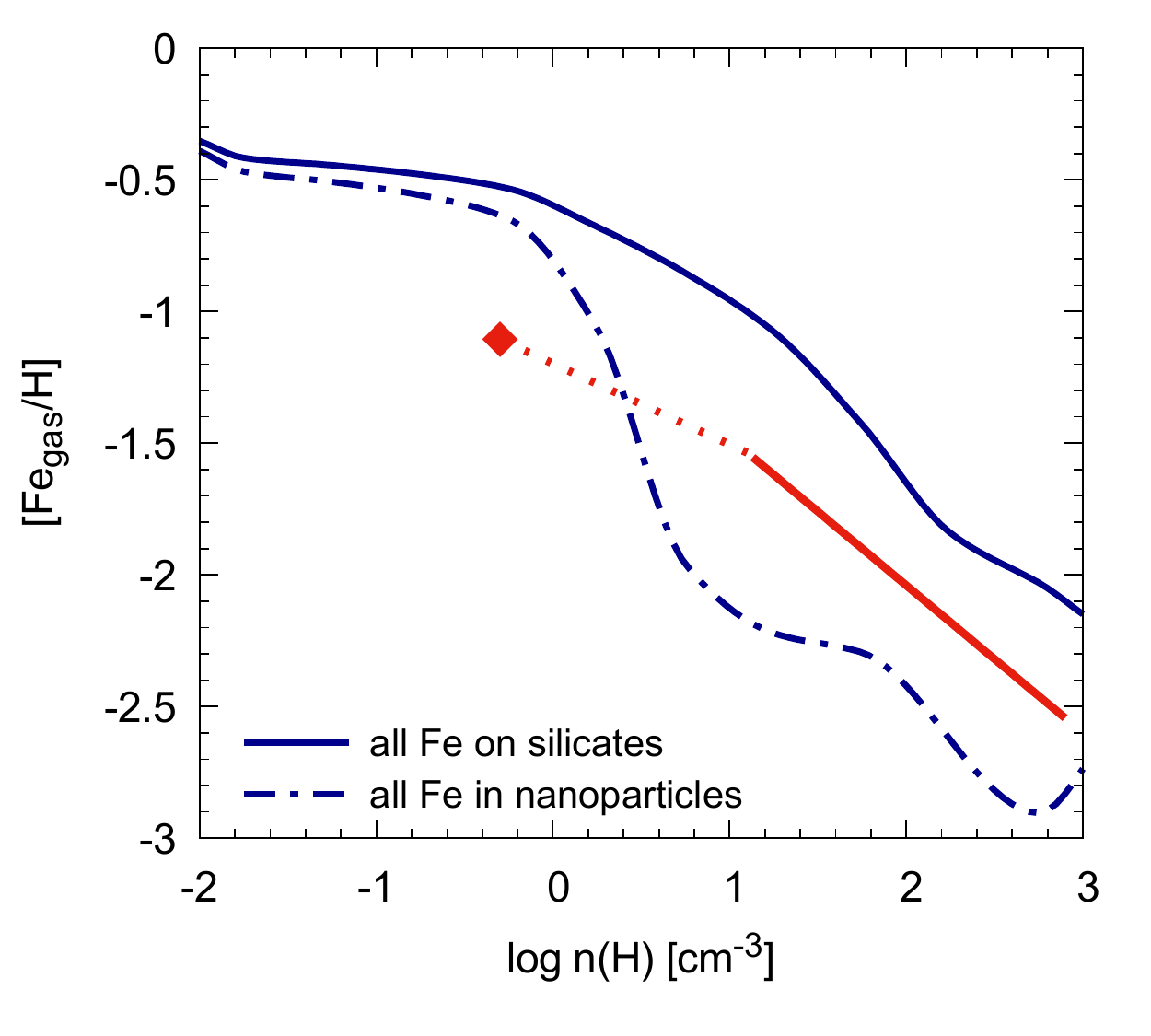}
\caption{Relations between mean \feh{} and gas density derived from the final spatial \feh{} distributions for two test simulations: (i) a case without iron nanoparticles, in which iron accretes on the silicate grain population (solid line) and (ii) a case of pure Mg-rich silicates with all solid iron residing in the form of metallic nanoparticles (dashed-dotted line). Only calculations which account for Coulomb focusing are shown. }
\label{fig:FeH-dens-rel-s}
\end{figure}

We also test two alternatives for iron-containing dust component: (i) a case without nanoparticles, in which iron accretes on the silicate grain population described in Sect.~\ref{sec:Si} and (ii) a case with pure Mg-rich silicates with all iron in solid form residing in nanoparticles (Fig.~\ref{fig:FeH-dens-rel-s}). Both models assume the same $m_{\rm cl}$ as for silicate grains. Model (i) fails to achieve the observed high Fe depletion level in the CNM which is explained by a lower total surface area and, consequently, an accretion rate compared to the reference model with nanoparticles. The amount of depleted Fe is limited by the timescales of formation of GMCs, which in this case are shorter than the accretion timescales. Model (ii) can provide sufficiently fast growth and reproduce both the observed level of Fe depletion in the CNM and the slope of \feh--\nh{} relation, but it overpredicts \feh{} in the WNM. During their residence in the diffuse gas, grains are too efficiently destroyed, so that \feh{} remains on the level of $-0.5$. 

Model (ii), in which all solid iron is in free-flying nanoparticles, can match the \feh{} value in the WNM, if a longer timescale of destruction $\tau_{\rm dest}^{\rm diff}=1.1\,$Gyr is adopted. This value seems unrealistically high, since it is unlikely that iron nanoparticles are destroyed 5 times less efficiently than silicate grains, given that postshock velocities of iron grains are 15\%-35\% higher than equivalent silicate grain velocities \citep{Jones:1994p1037}. 

\begin{figure}[t]
\includegraphics[scale=0.69]{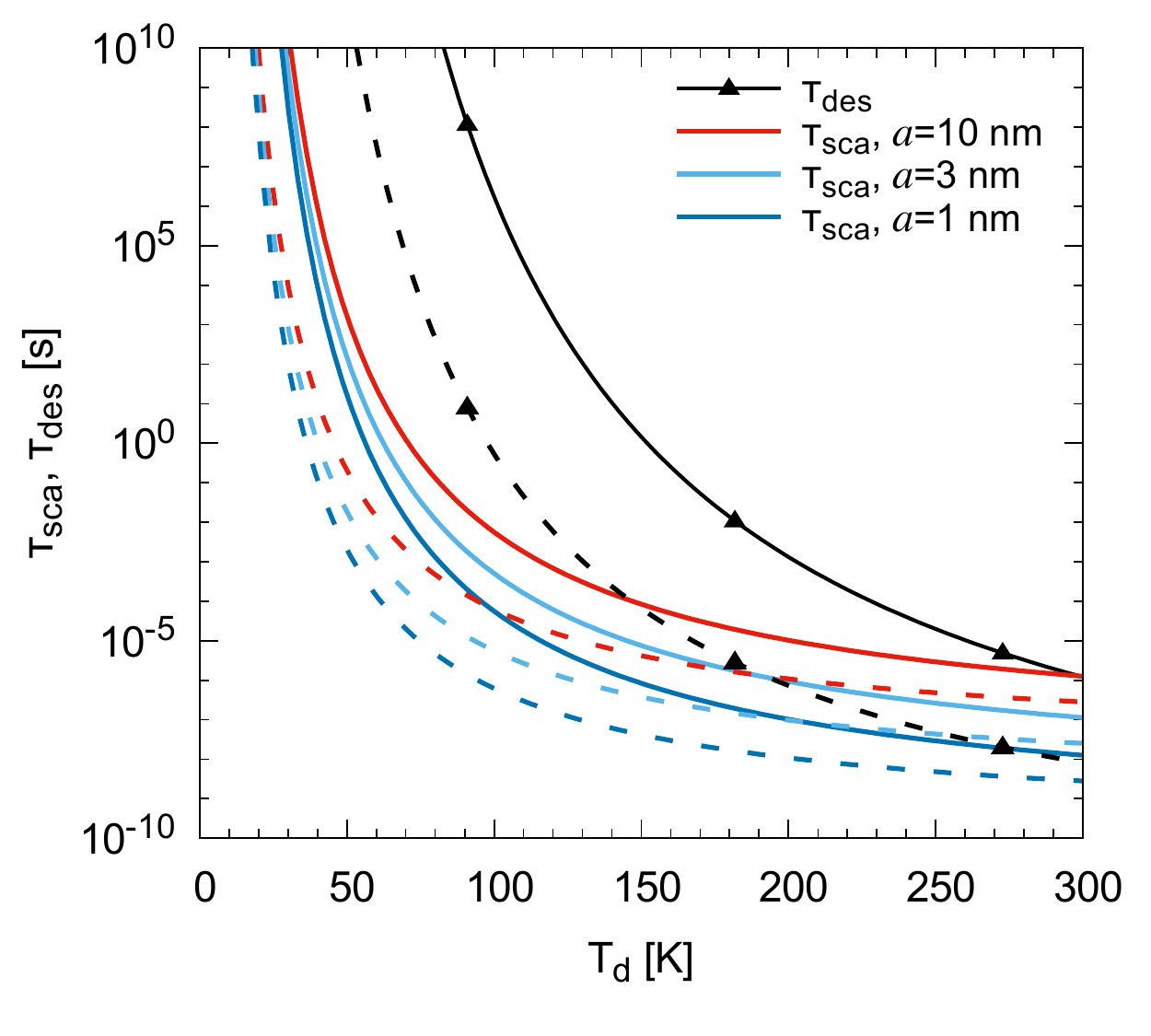}
\caption{Scanning timescales $\tau_{\rm sca}$ for grains with radii 10, 3 and 1 nm (red, light blue and dark blue lines, respectively) as a function of the grain temperature. Solid and dashed lines depict the timescales for $E_b/k_B=4200\,$K for Fe and $2700\,$K for Si, respectively. Desorption timescales $\tau_{\rm des}$ are shown with black solid lines with filled triangles.}
\label{fig:Timescales}
\end{figure}

\section{How do grains grow?}\label{sec:StochHeat}
 In this section we consider physical processes on the grain surface that may enable incorporation of new atoms into the solid.    Chemical kinetics of the surface reactions leading to the growth of species of interests at low temperatures is yet to be understood and studied experimentally. Majority of experimental work on silicate condensation is conducted at high temperatures ($\sim 1000\,$K), with only a few recent experiments so far that considered formation of silicate materials at temperatures below $15\,$K  \citep{Krasnokutski:2014bi, Rouille:2014gi, Henning:2017ta}. They find no activation energy barrier for the chemical reaction of low-temperature silicate formation. It is however not clear how this reaction proceeds in the intermediate temperatures. Nucleation and growth of Fe nanoclusters and deposition of Fe metallic films on various surfaces have been extensively studied around the room temperature and above \citep[e.g.,][and references therein]{Wastlbauer:2007fs, Lubben:2011bi}. Temperatures of iron grains in the diffuse medium are colder for most of time, but they raise drastically due to stochastic heating by UV photon absorptions \citep[][]{Draine:1985ei}. Stochastic heating of dust grains probably plays an important role in the formation of distinct silicate and carbon grain populations \citep{SerraDiazCano:2008p588, Draine:2009p6616}, but the unknown binding energies for the species of interests (Si, Mg, Fe) on the surface of interest (silicates and metallic iron) are still the main hindrance for modelling these processes in detail. We consider the timescales of relevant processes and their possible roles in the dust growth in the diffuse ISM.
 
\begin{figure}[t]
\includegraphics[scale=0.69]{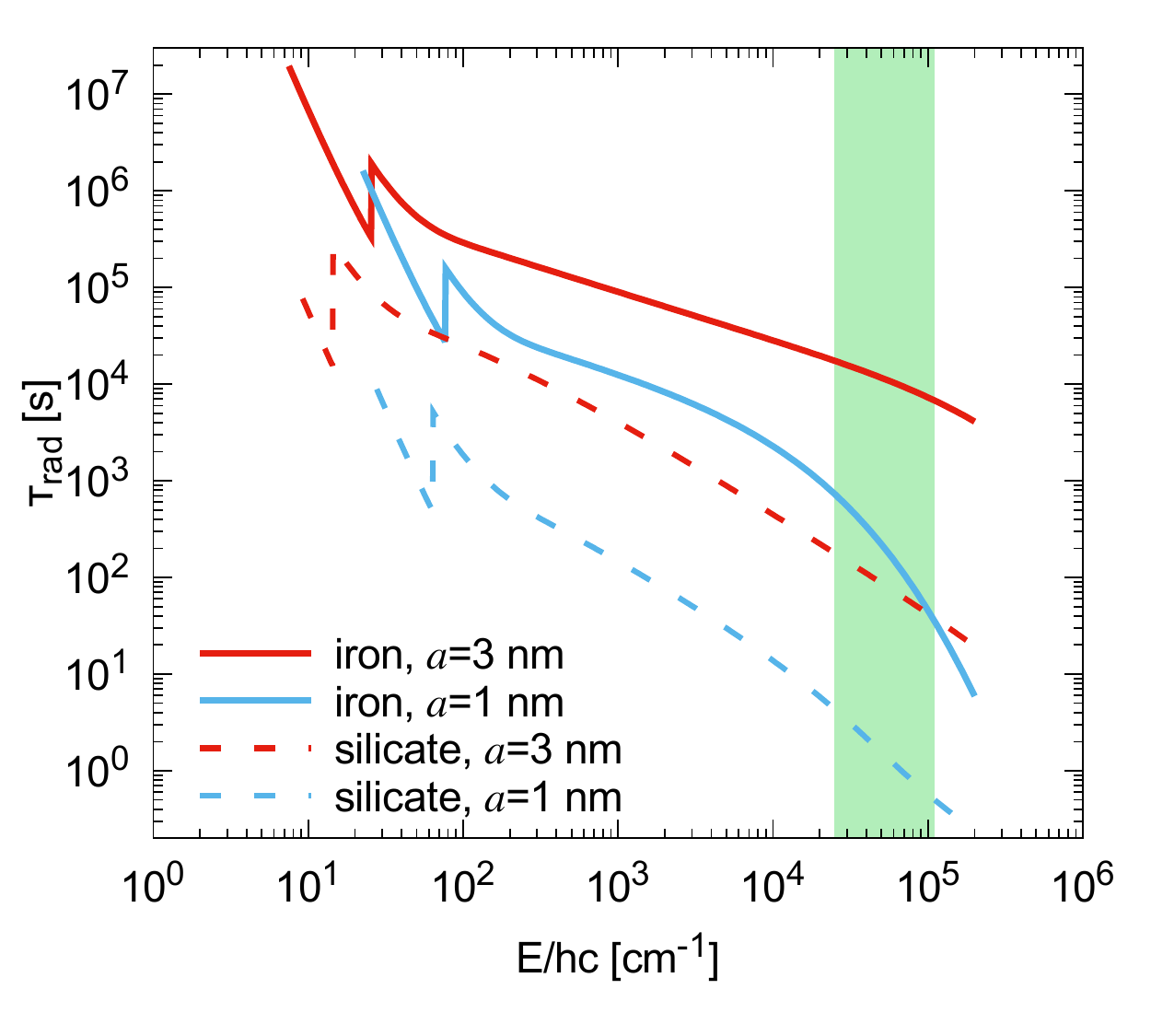}
\caption{Radiative cooling time $\tau_{\rm rad}$ as a function of vibrational energy $E$ for iron grains with radii 3 and 1 nm (solid red and blue lines, respectively). The same for silicate grains is shown with dashed lines for comparison. Rectangle indicates the UV wavelength range of the interstellar radiation field.}
\label{fig:tau-rad}
\end{figure}

 An impinging atom is initially bound by weak van der Waals forces to the surface. Our main assumptions are  that (1) the surface has active sites, i.e. sites with dangling bonds and high binding energy and (2)  the growth occurs when an adsorbed atom reaches an active site. Surface diffusion enables the transport of adatoms to the sites with high binding energy. The time to scan the entire surface by this process is 
\begin{equation}
\tau_{\rm sca}^{-1} \approx N^{-1}_{\rm s} \nu_0 \exp{(-E_{\rm d}/k_{\rm B} T_{\rm d} )},
\label{eq:tau_sca}
\end{equation}
where $E_{\rm d}$ is the diffusion energy,  $T_{\rm d}$ is the dust temperature, and $N_{\rm s}$ is the number of sites of the grain; $N_{\rm s} = 4 \pi a^2 n_s$, where $n_s\approx 1.5\times10^{15} \cms$ is the surface density of physisorption sites \citep{Hasegawa:1992p3930}. Note that for $a=1$nm $N_{\rm s}$ is only 188, while it is $10^6$  for a ``classical'' grain of $0.1\mum$. The diffusion energy is poorly known. Experimental work suggests that it is a fraction of the binding energy $E_{\rm b}$ of the adatom to the surface; we adopt $E_{\rm d}=0.3 E_{\rm b}$ here \citep{Hasegawa:1992p3930}.

Thermal desorption is the process that can remove an adsorbed atom from the grain surface before it reaches an active site. The timescale of thermal desorption is 
\begin{equation}
\tau_{\rm des}^{-1} \approx \nu_0 \exp{(-E_{\rm b}/k_{\rm B} T_{\rm d} )},
\label{eq:tau_des}
\end{equation}
where $\nu_0 \approx 10^{12}\, s^{-1}$ is the vibrational frequency of the sticking species.  

Both timescales $\tau_{\rm sca}$ and $\tau_{\rm des}$ are very sensitive to the binding energy. The value of $E_{\rm b}$ depends on many factors, including the chemical composition of the solid and its surface structure, in particular, on the number of unsatisfied bonds \citep{Cuppen:2017jd}. We take heuristic values for $E_{\rm b}/k_{\rm B}$ of 2700\,K for Si and 4200\,K for Fe from \cite{Hasegawa:1993go}. These values are similar to the physisoprtion energies estimated for water ice on silicate surface \citep{Stimpfl:2006is}. 

Figure~\ref{fig:Timescales} shows the timescales of desorption and scanning for Fe and Si atoms on grains with radii 1, 3 and 10\,nm. Both species exhibit a large range of dust temperatures, for which $\tau_{\rm sca} < \tau_{\rm des}$. In this range, adsorbed atoms can scan the entire grain surface and find the sites with the high binding energy before they are removed by thermal desorption. For $a=10$\,nm iron grains the scanning is faster than the thermal desorption for the dust temperatures up to $300\,$K and even higher for smaller sizes. In the following we consider temperatures of nanoparticles due to the stochastic heating.

\paragraph*{Role of stochastic heating.}
The temperature of small grains in the standard interstellar field fluctuates compared to the equilibrium value, because the cooling timescale for these grains is shorter than the mean time between consecutive absorptions of UV photons  \citep[][]{Draine:1985ei}. The maximum temperature $T_{\max}$ reached as a result of stochastic heating is determined  by absorption of the maximum photon energy of 13.6\,eV. For the smallest 1\,nm spherical iron grains considered in this work, the equilibrium temperature is 50\,K, while the actual  $T_{\rm d}$ can rise to $T_{\max}=270\,$K \citep{Fischera:2004gq}. Heating to these temperatures reduces the scanning time for Fe adatoms to $\lesssim10^{-7}$\,s (Fig.~\ref{fig:Timescales}). 
 The stochastic heating may therefore play an important role for the iron growth in the diffuse ISM. \textit{This holds only if the time $\tau_{\rm rad}$ needed for a grain heated to the temperature $T_{\rm d}$ to cool by emitting radiation is longer then the scanning timescale $\tau_{\rm sca}(T_{\rm d})$.}

In order to check this condition, we estimate the radiative cooling time for iron nanograins applying for them the formalism introduced by \cite{Draine:2001p4105} for silicate and PAH particles. \cite{Draine:2001p4105} represent a small grain with $N_{\rm a}$ atoms as a vibrational system with $3 N_{\rm a}-6$ degrees of freedom. The cooling time $\tau_{\rm rad}$ for a grain with vibrational energy $E_u$ gained by the absorption of an UV photon ($E_u=h\nu$) is then determined by its absorption coefficient $Q_{\rm abs}$ 
\begin{equation}
\tau_{\rm rad} (E_u) \approx \left[ \frac{1}{E_u} \frac{8 \pi^2 a^2}{h^3 c^2} {\int_0^{E_u} \frac{E^3 Q_{\rm abs}(E)}{\exp(E/k T_{u})-1} dE} \right],
\label{eq:tauradcool}
\end{equation}
where $T_{u}$ is the temperature of the vibrational system with energy $E_u$ in the notations of  \cite{Draine:2001p4105}. It corresponds to the grain temperature $T_{\rm d}$ in our notations. The calculation of $T_{u}$ and adopted data are briefly outlined in the Appendix~\ref{sec:Stochastics}.

The cooling timescales calculated from Eq.~(\ref{eq:tauradcool}) for iron and silicate grains are shown in Fig.~\ref{fig:Timescales} as a function of the absorbed photon energy. We show $\tau_{\rm rad} (E)$ for the grain sizes of 1 and 3\,nm. The lowest values of $\tau_{\rm rad}$ are derived for the maximum photon energy of 13.6\,eV of the interstellar UV field and equal to 30 and 7000\,s for 1 and 3\,nm sized iron grains. Silicate particles have lower values of $\tau_{\rm rad}$, 0.5 and 38\,s, respectively. This is the consequence of the higher opacities of silicates at long wavelenths ($\lambda \gtrsim 1\mum$) compared to iron grains.

The dust temperature rise from the UV absorption and consequently the reduction of $\tau_{\rm sca} $ decreases rapidly with the particle size. For example, for 3\,nm-sized iron grains $T_{\max}=70\,$K; the corresponding value of $\tau_{\rm sca}(70\,{\rm K}) = 0.11$\,s is significantly longer than $3\times10^{-8}$\,s for 1\,nm-sized particle, but is still shorter than the cooling time $\tau_{\rm rad}=7000$\,s. It is the particles of this size and smaller that have the largest total surface area and a higher fraction of negatively charged grains, therefore they are responsible for most of iron dust production in our simulations. The role of stochastic heating on the growth is expected to diminish for larger particles.

We estimate the mean time between UV photon absorptions $\tau_{\rm abs}$  to be equal to 1.2 and 35.5\,days for iron particle sizes of 3 and 1\,nm, correspondingly. Comparing these times to the time for radiative cooling, we find that small iron grains remain cool for most of time, in agreement with other studies of thermal behaviour of metallic grains \citep{Tabak:1987ke, Fischera:2004gq, Hensley:2017bc}. It is thus more likely that Fe atoms are accreted on cool grains and remain on the surface, since the probability of sticking is higher for lower $T_{\rm d}$. In the proposed picture of the dust growth, an accreted species can be delivered to an active site during the next UV absorption. The total number of such absorptions are $3\times 10^9$ and $10^8$ for 3 and 1\,nm sized grains. It is estimated for the mean residence time of grains in the cold medium inferred from high resolution hydrodynamic simulations and equal to 10\,Myr \citep{Peters:2017bl}. These numbers of absorptions are orders of magnitude higher than the number of Fe atoms accreted by iron grains. We therefore conclude that the stochastic heating may indeed assist in the growth of small iron grains in the cold ISM.

\section{Discussion}\label{sec:Discussion}
One-zone dust evolution models frequently assume that dust growth takes place in molecular clouds, because their high densities result in short timescales of accretion onto existing grains \citep[e.g.,][]{Dwek:1998p67, Hirashita:2000p462, Zhukovska:2008bw}. The main difficulty with silicate dust growth in molecular clouds, as summarised by \cite{2011A&A...530A..44J} and \cite{Ferrara:2016hma}, is that under such conditions grains accrete not only silicate-forming elements, but also abundant C-containing species (CO, CO$_2$ and other molecules) and H$_2$O that cover dust surfaces with ice mantles. Mantles have very different optical properties from the refractory dust materials required by the observed extinction features. 
 Moreover, ice mantles are weakly bound to the grain surfaces and rapidly evaporate as grains return to the diffuse medium \citep{Cuppen:2007p6467}. Thus, volatile mantles cannot replenish diffuse gas with dust upon disruption of molecular clouds and explain element depletion patterns and dust masses in galaxies. While refractory carbonaceous materials may be formed upon irradiation of ice mantles by UV photons \citep{Jenniskens:1993p2249}, explaining formation of amorphous silicates in molecular clouds remains problematic. 

The models for silicate and iron dust evolution presented in this work show that efficient growth of these dust components commences in the diffuse gas, specifically in its cold neutral component. The rate of dust production by this mechanism is 20 times higher than the total rate of dust injection by stars \citepalias{Zhukovska:2016jh}. Previous works found that the accretion timescale for grains with a ``classical'' radius of $0.1\mum$  is long compared to the residence time in the diffuse phase \citep{Ferrara:2016hma}. This discrepancy is solved, if we consider a grain size distribution including small grains. Because a fraction of small grains are negatively charged and species of interest are ionised (Fe, Si) in the CNM \citep{Weingartner:1999p6573, Yan:2004p865}, the collision rates of ions with grains are enhanced by Coulomb focusing. Assuming a MRN size distribution with a lower grain size of 4\,nm for silicate grains and accounting for the enhanced collision rate, we find the timescales of accretion in the CNM that are shorter than the residence time in this phase.  For iron grains, we need to extend the grain size distribution down to 1\,nm to explain the higher depletion of gas-phase Fe compared to Si. 

Temperature fluctuations of small grains due to the heating by interstellar UV photons may be vital for the dust growth, as  shown in Sect.~\ref{sec:StochHeat}.  The increase in the grain temperature drastically reduces the scanning time for an adsorbed species permitting it to reach a site with high binding energy before this species is removed from the surface by thermal desorption.

Dust growth in the diffuse ISM allows to explain reformation of distinct populations of silicate and carbonaceous grains, in contrast to the mantles grown in molecular clouds. Irradiation of grains by the UV photons plays a key role in the selective growth of silicate and carbonaceous grains as separate species \citep{Draine:1990p495, Draine:2009p6616}. It is based on the assumption that the binding energy of carbon atoms on silicate grain surface is smaller than that of Mg, Si, Fe and O, therefore C can be cleared from their surfaces by photo-desorption. Recent experimental work on the condensation of Si, O, Mg and Fe on the low-temperature substrate indicates that these species are indeed able to react without an activation energy barrier. They form siliceous material that remains stable at the room temperature and shows 10\mum{} spectral band which is very similar to the silicate absorption feature observed in the ISM \citep{ Krasnokutski:2014bi, Rouille:2014gi}. Recent experiments have demonstrated that even in a situation where one starts with mixed precursors one gets separate silicate and carbonaceous particles \citep{Henning:2017ta}. 

In the standard interstellar UV radiation field, Coulomb interactions enhance collision rates and enables dust growth by accretion at densities as low as  $\nh \gtrsim 1\cmc$. It is, however, noteworthy that physical conditions for dust growth in early star-forming galaxies are likely very different. In regions with strong radiation field, dust grains are charged positively and Coulomb repulsion may hamper dust growth \citep{Ferrara:2016hma}. Moreover, the ISM of galaxies forming stars at a 100 times higher rate than our own is expected to be highly turbulent. To discern whether dust growth is a viable source in the early Universe requires better understanding of grain charging and matter cycle under extreme ISM conditions and is subject of our future studies.

\section{Conclusions}\label{sec:Conclusions}
The interstellar abundance of refractory elements indicate substantial depletion that increases with gas density. Our new dust evolution model based on hydrodynamic simulations of the lifecycle of GMCs established that the observed relation between the mean gas-phase Si abundance \sih{} and the local gas density \nh{} is driven by a combination of selective silicate dust growth by accretion and efficient destruction in the diffuse ISM \citepalias{Zhukovska:2016jh}. The model naturally includes different residence times of grains in  interstellar phases and dependence of dust growth on the local density and temperature of the gas. In this work,  we reinforce this conclusion with a simple analytic model  for the Si depletion assuming a steady state between destruction and production of dust by stellar sources and by accretion in the ISM. We demonstrate that dust production by stars cannot reproduce the mean $\sih-\nh$ relation even for optimistic assumptions for the condensation efficiencies and absence of dust destruction in SN shocks. This analytic model illustrates how the dependence of accretion timescale on gas density determines the negative slope of the $\sih-\nh$ relation. 

We extend the framework developed in our previous work for silicates to include the evolution of iron dust, using the average $\feh - \nh$ relation inferred from observations as main constraint for the iron grain population. In order to explain a 0.9~dex lower value of  \feh{} in the CNM compared to the \sih{} value, we need a population of free-flying metallic nanoparticle with radii from 1 to 10\,nm. However, if all iron missing from the gas resides in nanoparticles and their destruction is similar to that of silicate grains, the model overpredicts the \feh{} in the WNM by 0.5~dex compared to the observed depletion. The slope of the \feh--\nh{} relation is steeper than the observed value. This discrepancy is solved, if we assume that a fraction of the depleted iron  of  70\% resides in the form of metallic inclusions inside of silicate grains, where it is protected from rapid destruction by interstellar shocks. Alternatively, the observed depletion trend can be described with a model, in which all depleted Fe resides in nanoparticles, with the condition that their destruction efficiency is 5 times lower than that of the silicate grain population. Such low destruction of metallic nanoparticles is presently difficult to reconcile with predictions by theoretical models for dust destruction in interstellar shocks.

Enhanced collision rates due to the electrostatic focusing in the CNM are crucial for both silicate and iron dust models to reproduce the slope of the observed depletion--density relations and the magnitude of depletions at high densities. Without taking the grain charges into account, the timescales of accretion are longer than the timescale of formation of GMCs and the resultant [X$_{\rm gas}$/H] values in the CNM are higher than the observed values. 

We consider the timescales of relevant physical processes for adsorbed Si and Fe atoms on the grain surfaces. A process, which is presently not included in the dust evolution models, but may be important for the dust growth is stochastic heating of small grains by UV photon absorption. We demonstrate that the timescale for radiative cooling of a grain after an absorption is longer than the timescale for scanning of the surface by an adsorbed species. The increase in the grain temperature drastically reduces the scanning time and enables delivery of adsorbed species to the sites with high binding energy. 

\acknowledgments

We thank anonymous referee for helpful comments that  improved the clarity of the paper. S.Z. acknowledges support by the Forschungsgemeinschaft through SPP 1573: “Physics of the Interstellar Medium”. We gratefully acknowledge the Max Planck Computing and Data Facility for providing their user support and computing time on the Odin and Hydra clusters. 

\appendix

\section{Density dependence of dust destruction}\label{sec:A1}
The quantity $m_{{\rm cl},j}$ is determined by the properties of the dust material and the structure of the blast wave
\begin{equation}
 m_{{\rm cl},j}(n_0) = \int_{\varv_0}^{\varv_f} \epsilon_j(\varv_s,n_0) \left| \frac{{\rm d}M_s(\varv_s,n_0)}{{\rm d}\varv_s} \right| {\rm d} \varv_s\,,
	\label{eq:MassCleared}
\end{equation} 
where $\varv_0$ and $\varv_f$ are the initial and final velocities of the SNR expanding into an ambient medium of density $n_0$, respectively, $\left| \frac{{\rm d}M_s(\varv_s,n_0)}{{\rm d}\varv_s} \right| d\varv_s$ is the mass of gas swept up by a shock with velocity in the range of $[\varv_s,\varv_s+d\varv_s]$, and $\epsilon_j$ is the efficiency of dust destruction  in a SN shock with expansion velocity $\varv_s$.  We calculate $\left| \frac{{\rm d}M_s(v_s,n_0)}{{\rm d}v_s} \right|$ using an analytical solution for the SNR evolution expanding in a homogeneous medium from \cite{McKee:1989p1030}. The solution combines the adiabatic expansion and pressure-driven snow plow stages. We adopt an expression for $\epsilon_j(\varv_s,n_0)$ for silicate dust calculated by \cite{Jones:1996p6593}, which is available only for the ambient density $n_0=0.25\cmc$. Although the efficiency of dust destruction in SN shock depends on the density of ambient medium  \citep{Jones:1994p1037, Jones:1996p6593, Nozawa:2006p1022}, it should not significantly vary in the density range of  0.2--1\cmc,  which encompasses 70\% of the diffuse gas in our simulations. 
We derive the following quadratic fitting formula for this integral for metallicity $Z=0.014$
\begin{equation}
m_{{\rm cl},j}(n_0) = a (\log n_0)^2 + b \log n_0 + c,
\label{eq:A1}
\end{equation}
where $a=22$, $b=-305.85$, and $c=1438.93$ for silicate dust. This formula is accurate within 1\% w.r.t. the numerical integration of Eq.~(\ref{eq:A1}).

\section{Stochastic heating of nanoparticles}\label{sec:Stochastics}
In order to calculate the radiative cooling time for an iron grain upon absorption of a photon (Eq.~(\ref{eq:tauradcool})), we apply a statistical-mechanical description of the emission process \citep{Draine:2001p4105}. A grain with $N_{\rm a}$ atoms is approximated  as a vibrational system with $N_m = 3N_{\rm a} - 6$ degrees of freedom. The main assumptions of this approach are that (1) the energy of the absorbed photon $h\nu$ is distributed ergodically among these degrees of freedom before any infrared emission and (2) absorption coefficient is independent of the degree of excitation. An assumption that the vibrational modes of the grain are approximated by harmonic oscillators  allows us to calculate the temperature $T_u$ of grains in the excited state $E_u=h\nu$ from the equation $\overline{E}(T)=E_u$, where $\overline{E}$ is the expectation value for the vibrational energy:
\begin{equation}
\overline{E}(T) = \sum_{j=1}^{N_m} \frac{\hslash \omega_j}{\exp(\hslash \omega_j/kT) - 1},
\end{equation}
where $\omega_j$ is the fundamental frequency of the mode $j$. We adopt an $n$-dimensional Debye spectrum for the fundamental modes for silicate and iron grains described respectively in \cite{Draine:2001p4105} and \cite{Hensley:2017bc}.

For metallic iron, we adopt $Q_{\rm abs}(a,\lambda)$ from \cite{Fischera:2004gq}, who calculated the optical properties of spherical pure iron grains using Mie-theory for wavelentghs $2\pi a/\lambda>0.1$ and dipole approximation at longer wavelengths. Optical properties for astronomical silicates are taken for silicate grains \citep{Draine:1984p459, Laor:1993p1280}.

\bibliography{AllReferences}

\end{document}